\documentclass[preprint,aps]{revtex4}
\usepackage{epsfig}
\begin{document}

\title{Completeness of non-normalizable modes}

\author{Philip D. Mannheim}\email{philip.mannheim@uconn.edu}
\author{Ionel Simbotin}\email{simbotin@phys.uconn.edu}
\affiliation{Department of Physics,
University of Connecticut, Storrs, CT 06269}

\date{July 13, 2006}

\begin{abstract}

We establish the completeness of some characteristic sets of 
non-normalizable modes by constructing fully localized square steps out of
them, with each such construction expressly displaying the Gibbs
phenomenon associated with trying to use a complete basis of modes to fit
functions with discontinuous edges. As well as being of interest in and
of itself, our study is also of interest to the recently introduced large
extra dimension brane-localized gravity program of Randall and Sundrum,
since the particular non-normalizable mode bases that we consider
(specifically the irregular Bessel functions and the associated Legendre
functions of the second kind) are associated with the tensor
gravitational fluctuations which occur in those specific brane worlds in
which the embedding of a maximally four-symmetric brane in a
five-dimensional anti-de Sitter bulk leads to a warp factor which is
divergent. Since the brane-world massless four-dimensional graviton has a
divergent wave function in these particular cases, its resulting lack of
normalizability is thus not seen to be any impediment to its belonging to
a complete basis of modes, and consequently its lack of normalizability
should not be seen as a criterion for not including it in the spectrum of
observable modes. Moreover, because the divergent modes we consider
form complete bases, we can even construct propagators out of them in
which these modes appear as poles with residues which are expressly
finite. Thus even though normalizable modes appear in propagators with
residues which are given as their finite normalization constants,
non-normalizable modes can just as equally appear in propagators with
finite residues too -- it is just that such residues will not be
associated with bilinear integrals of the modes.

\end{abstract}

\maketitle

\section{Introduction}

In constructing complete bases of mode solutions to wave equations it is
very convenient to work with modes which are normalizable since they obey
a closure relation. Specifically, if one has some complete
orthonormal basis of modes $f_m(w)$ with eigenvalues labelled by $m$ and
orthonormality relation 
\begin{equation}
\int^{\infty}_{-\infty}dw e^{-2A(w)}f_{m}(w)f_{m^{\prime}}(w)
=\delta_{m,m^{\prime}}
\label{1}
\end{equation}
where $e^{-2A(w)}$ is an appropriate normalization measure, the
completeness of the basis will then require that any localized
function be expandable in terms of the basis modes as 
\begin{equation}
\psi(w)=\sum_{m} a_mf_m(w)
\label{2}
\end{equation}
with coefficients which are given as
\begin{equation}
a_m=\int^{\infty}_{-\infty}dw e^{-2A(w)}\psi(w)f_{m}(w)~~.
\label{3}
\end{equation}
With insertion of these coefficients back into Eq. (\ref{2}) yielding
\begin{eqnarray}
\psi(w)&=&\int^{\infty}_{-\infty}dw^{\prime}
\delta(w-w^{\prime})\psi(w^{\prime})
\nonumber \\
&=&
\sum_{m}\int^{\infty}_{-\infty}dw^{\prime}
e^{-2A(w^{\prime})}\psi(w^{\prime})f_m(w^{\prime})f_m(w)~~,
\label{4}
\end{eqnarray}
the arbitrariness of the choice of $\psi(w)$ will then require that the
basis modes obey a closure relation of the form 
\begin{equation}
\sum_{m} f_{m}(w^{\prime})f_m(w)
=e^{2A(w)}\delta(w-w^{\prime})~~.
\label{5}
\end{equation}
With Eq. (\ref{5}) being recognized as being a special case of Eq.
(\ref{2}) (viz. the expansion of the extremely localized $\delta(w)$ in a
complete basis of $f_m(w)$ with coefficients $a_m=f_m(0)$), the notions
of completeness and closure are often treated interchangeably in the
literature, with Eq. (\ref{5}) not only often being referred to as being a
completeness relation, but with it even being regarded as being an
essential requirement for a basis to be complete in the first place. 

It is the purpose of this paper to show that this need not in fact be the
case, and that modes can be complete even when they do not obey Eq.
(\ref{5}) at all. Indeed, the steps which lead from Eq. (\ref{1}) to Eq.
(\ref{5}) only hold when the basis is in fact one for which the integrals
on the left-hand side of Eq. (\ref{1}) do in fact exist. With both of
Eqs. (\ref{1}) and (\ref{5}) involving bilinear functions of the basis
modes, but with Eq. (\ref{2}) only being a linear function of the modes,
it is still possible for the summation in Eq. (\ref{2}) to be
well-defined even when the bilinear expressions which appear in Eqs.
(\ref{1}) and (\ref{5}) are not. Moreover, the wave equations for which
the $f_m(w)$ are the mode solutions are themselves only linear functions
of the $f_m(w)$, and it should thus be immaterial to the completeness
of their solutions as to whether or not bilinear integrals of the
modes exist. In general then completeness of a basis has to be understood
as being the requirement that for localized functions $\psi(w)$ there
exists an expansion of the form of Eq. (\ref{2}) with finite coefficients
$a_m$ regardless of whether or not the integrals on the left-hand side of
Eq. (\ref{1}) actually exist. Non-normalizable modes whose behavior is so
bad as to cause these bilinear integrals to diverge can still be complete
in the sense of Eq. (\ref{2}), with the $a_m$ coefficients being such as
to lead to total destructive interference between the $f_m(w)$ in the
regions where the $f_m(w)$ diverge. It is thus Eq. (\ref{2}) which has to
be recognized as being the general statement of completeness, and in this
paper we shall confirm this by explicitly constructing localized square
steps as sums over some characteristic bases of divergent modes. While
the existence or not of the normalization integrals of Eq. (\ref{1}) is
immaterial to a differential wave equation, if the solutions to the wave
equation are required to belong to a Hilbert space one can restrict to
square integrable functions alone, though otherwise there is no reason to
discard any non-normalizable solutions \cite{footnote0}. Since wave
equations in classical physics do not act in a Hilbert space, in
classical physics one is not free to discard
non-normalizable modes, and since classical physics wave equations play a
prominent role in classical gravity where they are associated with
classical gravitational fluctuations around classical gravity
backgrounds, it is to classical gravity that we shall look for examples
in which to test whether non-normalizable modes can be complete. 

\section{Wave equations for gravitational fluctuations}

The wave equations we shall explicitly explore are associated with the
recently introduced brane-localized gravity program of Randall and Sundrum
\cite{Randall1999a,Randall1999b}. As introduced, the brane gravity
program provides for the possibility that our four-dimensional universe
could be embedded in some infinitely-sized bulk space and yet not 
conflict with the fact that there is no apparent sign of any such
higher-dimensional bulk. Specifically, by taking the higher-dimensional
bulk to possess a very special geometry, viz. the five-dimensional
anti-de Sitter geometry $AdS_5$, and by taking our four-dimensional
universe to be a brane (viz. membrane) embedded in it, Randall and
Sundrum found that under certain circumstances it was then possible for
gravitational signals to localize around the brane and not penetrate very
far into the bulk, with $AdS_5$ acting as a sort of refractive medium
which rapidly attenuates any signals which try to propagate in it.
Within the Randall-Sundrum brane world there are six fully soluble set-ups
(technically $AdS_5$ bulks with embedded Minkowski, de Sitter or anti-de
Sitter branes each with either positive or negative tension $\lambda$ --
to be referred to as the $M_4^{\pm}$,
$dS_4^{\pm}$ and $AdS_4^{\pm}$ brane worlds in the following), with all
six of them having backgrounds which can be described by the generic
five-dimensional metric
\begin{equation}
ds^2=dw^2+e^{2A(|w|)}q_{\mu \nu}(x^{\lambda})dx^{\mu}dx^{ \nu}
\label{6}
\end{equation}
where the $w$-independent $q_{\mu\nu}$ is the four-dimensional metric and
the so-called warp factor $e^{2A(|w|)}$ is taken to be a function of $|w|$
where
$w$ is the fifth coordinate.  With the curvature of $AdS_5$ being taken to
be given as $-b^2$, in the various cases the explicit background metrics
are given as 
\begin{equation}
ds^2(M_4^{\pm})=dw^2+e^{-2\epsilon(\lambda)b|w|}[dx^2+dy^2+dz^2-dt^2]~~,
\label{7}
\end{equation}
\begin{equation}
ds^2(dS_4^{\pm})=dw^2+{H^2\over b^2}{\rm
sinh}^2\left[{\rm arcsinh}\left({b \over H}\right)-\epsilon(\lambda)b|w|
\right]
\left[e^{2Ht}(dx^2+dy^2+dz^2)-dt^2\right]~~,
\label{8}
\end{equation}
and 
\begin{equation}
ds^2(AdS_4^{\pm})=dw^2+{H^2\over b^2}{\rm
cosh}^2\left[{\rm arccosh}\left({b \over H}\right)-\epsilon(\lambda)b|w|
\right]
\left[dx^2+e^{2Hx}(dy^2+dz^2-dt^2)\right]~~,
\label{9}
\end{equation}
where $\epsilon(\lambda)$ is the sign of $\lambda$. (The $M_4^{\pm}$
background metrics are given in \cite{Randall1999a,Randall1999b}, the 
$dS_4^{\pm}$ background metrics are given in \cite{DeWolfe2000,Kim2000},
and the $AdS_4^{\pm}$ background metrics are given in
\cite{DeWolfe2000}.)

For the brane world the gravitational fluctuations around these
six backgrounds are most readily treated in the axial gauge where the
transverse-traceless tensor fluctuation modes $h^{^{TT}}_{\mu\nu}$ then 
all obey the generic wave equation (see e.g. \cite{Mannheim2005} where
full derivations and relevant citations are given)
\begin{equation}
\left[{\partial^2 \over \partial |w|^2}
-4\left({d A\over
d|w|}\right)^2
+e^{-2A}\tilde{\nabla}_{\alpha}\tilde{\nabla}^{\alpha}
\right]h^{^{TT}}_{\mu\nu} =0~~,
\label{10}
\end{equation}
as subject to the constraint (technically the Israel junction condition)
\begin{equation}
\delta(w)\left[{\partial \over
\partial |w|}-2{d A\over
d|w|}\right]h^{^{TT}}_{\mu\nu}=0
\label{11}
\end{equation}
at a brane which is located at $w=0$. In Eq. (\ref{10}) the tildas in
$\tilde{\nabla}_{\alpha}\tilde{\nabla}^{\alpha}$ indicate that these
particular covariant derivatives are to be evaluated in the geometry
associated with the four-dimensional $q_{\mu\nu}$. And with the
four-dimensional sector of the theory being separable according to 
\begin{equation}
[\tilde{\nabla}_{\alpha}\tilde{\nabla}^{\alpha} -2kH^2]h^{^{TT}}_{\mu\nu}=
m^2h^{^{TT}}_{\mu\nu}~~,
\label{12}
\end{equation}
as defined here so that tensor fluctuations with $m^2=0$  propagate
on the  appropriate $dS_4$, $M_4$ or $AdS_4$ lightcones ($k=1,0,-1$
respectively), a separation of the modes into the form
$h_{\mu\nu}^{^{TT}}=f_m(|w|)e_{\mu\nu}(x^{\lambda},m)$ then requires that
the $f_m(|w|)$ obey 
\begin{equation}
\left[{d^2 \over d |w|^2}
-4\left({d A\over
d|w|}\right)^2 
-2\left({d^2 A\over
d|w|^2}\right)
+e^{-2A}m^2
\right]f_m(|w|)=0
\label{13}
\end{equation}
(in each of the six background cases of interest to us the identity
$d^2A/d|w|^2= -kH^2e^{-2A}$ holds), as subject to the constraint 
\begin{equation}
\delta(w)\left[{d \over
d|w|}-2{d A\over
d|w|}\right]f_m(|w|)=0~~. 
\label{14}
\end{equation}
Our task is thus to explore the completeness of solutions to Eqs.
(\ref{13}) and (\ref{14}), and a reader unfamiliar with the physics
of the brane world can start at this point as none of the analysis which
ensues will depend on how Eqs. (\ref{13}) and (\ref{14}) were first
arrived at. What will matter in the following is only that these equations
admit of exact solutions, solutions whose large $|w|$ behavior can
then explicitly be monitored.

Before actually identifying explicit solutions to Eqs. (\ref{13}) and
(\ref{14}) for the specific choices of $A$ and $\epsilon(\lambda)$ of
interest, we note that via manipulation of Eq. (\ref{13}) we find that 
every pair of its solutions have to obey 
\begin{eqnarray}
&&e^{-2A}(m_1^2-m_2^2)f_{m_1}f_{m_2}
\nonumber \\
&&\phantom{=}=
{ d \over d |w|}\left[f_{m_1}\left({d \over d|w|}-2{dA \over d|w|}\right)
f_{m_2} -f_{m_2}\left({d \over d|w|} -2{dA \over
d|w|}\right)f_{m_1}\right]~~,
\label{15}
\end{eqnarray}
which with Eq. (\ref{14}) then requires the modes to obey
\begin{eqnarray}
&&(m_1^2-m_2^2)\int_0^{\infty}d|w|e^{-2A}f_{m_1}f_{m_2}
\nonumber \\
&&\phantom{=}=
\lim_{|w| \rightarrow \infty}
\left[f_{m_1}\left({d \over d|w|}-2{dA \over d|w|}\right)f_{m_2}
-f_{m_2}\left({d \over d|w|}-2{dA \over d|w|}\right)f_{m_1}\right]~~.
\label{16}
\end{eqnarray}
Orthogonality of modes with different separation constants is thus
achieved when the modes are well-enough behaved at $|w|=\infty$
to cause the right-hand side of Eq. (\ref{16}) to vanish (with the
orthogonality measure then being precisely the one we introduced in Eq.
(\ref{1})), with modes which diverge badly enough at infinity causing the
integral on the left-hand side to not exist. While one could now proceed
to determine the mode solutions and identify for which particular ones the
integral on the left-hand side of Eq. (\ref{16}) converges or diverges,
before doing so it is instructive to recall that via a sequence of
transformations it is possible to bring Eq. (\ref{13}) to a more familiar
form. Specifically, if we change variables from $w$ to $z$ by setting
$dz=e^{-A(w)}dw$ and define $f_m=e^{A(z)/2}\hat{f}_m$, the $\hat{f}_m$
will then obey
\cite{Randall1999b} 
\begin{equation}
\left[-\frac{d^2}{d z^2} +{9\over
4}\left({dA \over dz}\right)^2+{3\over 2}{d^2A
\over dz^2}-m^2\right]\hat{f}_m=0~~,
\label{17}
\end{equation}
while at the same time the normalization integral will change as
\begin{equation}
\int_0^{\infty}d|w|e^{-2A}f_{m_1}(|w|)f_{m_2}(|w|) \rightarrow
\int_{z[0]}^{z[\infty]}dz\hat{f}_{m_1}(z)\hat{f}_{m_2}(z)~~.
\label{18}
\end{equation}
While we thus recognize Eq. (\ref{17}) as being in the familiar form of a
one-dimensional $\rm{Schr\ddot{o}dinger}$ equation and Eq. (\ref{18}) as
being in the form of its conventional quantum-mechanical normalization
integral, nonetheless, as noted above, since in the cases which are of
interest to us here we are not requiring the $\hat{f}_m$ modes to belong
to a Hilbert space, we should not discard the non-normalizable
solutions to Eq. (\ref{17}) \cite{footnote1}. And having now recognized
the rationale for not discarding non-normalizable solutions, we return
to Eqs. (\ref{13}) and (\ref{14}) to actually find and then explore
them.  

\section{Completeness tests for the Minkowski brane cases}

\subsection{Positive tension case}

For the $M_4^+$ case where $A=-b|w|$ the solutions to Eq.
(\ref{13}) are readily obtained by setting $y=me^{b|w|}/b$ as this
transformation brings Eq. (\ref{13}) to the Bessel equation form
\begin{equation}
\left[{d^2 \over dy^2}+{1 \over y}{d \over dy} +1 -{4 \over
y^2}\right]f_m(y)= 0~~.
\label{19}
\end{equation}
Mode solutions with any positive $m^2$ are thus given by
\begin{equation}
f_m(y)=\alpha_m J_{2}(y)+\beta_m Y_{2}(y)
\label{20}
\end{equation}
where $\alpha_m$ and $\beta_m$ are $y$-independent coefficients,  
with those solutions with $m^2=0$ being given directly from Eq. (\ref{13})
as 
\begin{equation}
f_0(y)=\alpha_0 e^{-2b|w|}+\beta_0 e^{2b|w|}~~.
\label{21}
\end{equation}
To satisfy the junction condition of Eq. (\ref{14}) then requires that
the various mode coefficients obey
\begin{equation}
\alpha_m J_1(m/b)+\beta_m
Y_1(m/b)=0~~,~~\beta_0=0~~,
\label{22}
\end{equation}
with the continuum of $m^2>0$ modes thus satisfying the junction condition
via an interplay of the two types of Bessel function, and the $m^2=0$ mode
$f_0(y)=\alpha_0 e^{-2b|w|}$ satisfying it all on its own. In the brane
world the $m^2>0$ modes are known as the KK (Kaluza-Klein) modes, while
the $m^2=0$ mode serves as a massless graviton. At large $y$ these
solutions behave as 
\begin{equation}
f_m\rightarrow \left({2 \over \pi y}\right)^{1/2}\left[\alpha_m
{\rm cos}\left(y-5\pi/4\right)+\beta_m
{\rm sin}\left(y-5\pi/4\right)\right]~~,~~f_0 \rightarrow
{\alpha_0m^2 \over b^2y^2}~~.
\label{23}
\end{equation}
With all of these modes having wave functions which fall very fast in
$|w|$ as we go away from the brane, the gravitational fluctuation modes
are thus localized around it, this being the key result of
\cite{Randall1999b}. With the measure of the normalization integral being
rewriteable as 
\begin{equation}
\int_0^{\infty}d|w|e^{2b|w|}=b
\int_{1/b}^{\infty}dx x
\label{24}
\end{equation}
on setting $x=e^{b|w|}/b$, we see that the massless graviton wave
function is bound state normalizable and that the KK modes possess the
same continuum normalization as flat space Bessel functions.
Consequently, the totality of massless graviton plus KK continuum modes
is complete in exactly the same way as plane waves, with both of Eqs.
(\ref{1}) and (\ref{5}) being satisfied (the summation in Eq. (\ref{5})
is understood to contain both discrete and continuous indices). While we
thus see that there is no need to perform any explicit completeness test
for the modes of $M_4^+$ as everything is standard, a quite different
situation will emerge when we consider $M_4^-$.

\subsection{Negative tension case}

For the $M_4^-$ case where $A=+b|w|$, the $m^2>0$ and the
$m^2=0$ solutions to Eq. (\ref{13}) are given by  
\begin{equation}
f_m(y)=\alpha_m J_{2}(y)+\beta_m Y_{2}(y)~~,
\label{25}
\end{equation}
and 
\begin{equation}
f_0(y)=\alpha_0 e^{-2b|w|}+\beta_0 e^{2b|w|}~~,
\label{26}
\end{equation}
where now $y=me^{-b|w|}/b$, while to satisfy the junction
condition of Eq. (\ref{14}) this time requires 
\begin{equation}
\alpha_m J_1(m/b)+\beta_m
Y_1(m/b)=0~~,~~\alpha_0=0~~.
\label{27}
\end{equation}
Unlike the $M_4^+$ case this time $y$ goes to zero as $|w|$ goes to
infinity, with large $|w|$ asymptotics now being controlled by the
behavior of Bessel functions at small argument rather than large, with
the solutions behaving at small
$y$ as 
\begin{equation}
f_m\rightarrow {\alpha_m y^2 \over 8}-{4\beta_m
\over \pi y^2}~~,~~f_0 \rightarrow
{\beta_0m^2 \over b^2y^2}
\label{28}
\end{equation}
(the $Y_2(y)$ behave irregularly at small argument). With the measure of
the normalization integral now being given as 
\begin{equation}
\int_0^{\infty}d|w|e^{-2b|w|}=b
\int_{0}^{1/b}dx x
\label{29}
\end{equation}
on setting set $x=e^{-b|w|}/b$, this time we see that it is only the
$J_2(y)$ modes which are normalizable, and that the massless graviton
wave function and all the $Y_2(y)$ modes are not only non-normalizable,
they diverge far too violently to even be plane wave normalizable. In
order to be able to satisfy the junction condition of Eq. (\ref{27}) with
normalizable modes alone, the convergent $J_2(y)$ modes would have to
satisfy Eq. (\ref{27}) all by themselves, with the modes then needing to
obey
$J_1(m/b)=0$. Solutions to this condition exist, and are given as the
zeroes, $j_i$, of the Bessel function $J_1$. This set of zeroes is 
discrete and infinite, with the normalizable modes of the $M_4^-$ brane
world then being given as modes with masses $m_i=bj_i$. Similarly, the
divergent $Y_2(y)$ modes can satisfy the junction condition all on their
own if their masses obey $m_i=by_i$ where the $y_i$ are the zeroes of the
Bessel function $Y_1$, to yield another infinite set of discrete modes.
With the divergent massless graviton mode with wave function
$\beta_0e^{2b|w|}$ also satisfying the junction condition on its own, we
thus recognize two classes of basis modes in the $M_4^-$ brane world, the
convergent $J_2(j_ie^{-b|w|})$, and the divergent $e^{2b|w|}$ and
$Y_2(y_ie^{-b|w|})$. And while our objective is to apply a completeness
test to the divergent mode basis, it will be instructive to actually
apply a completeness test to the convergent $M_4^-$ mode basis first.

\section{Completeness test for convergent $M_4^-$ modes}

To test for completeness of a basis we need to determine whether it is
possible to expand the typical localized square step $V_J=\hat{V}$,
$\alpha \leq e^{-b|w|}/b \leq \beta$, $V_J=0$ otherwise in terms of the
modes of this basis, viz. we seek to find a set of $V_m$ from which we can
reconstruct the square step according to
\begin{equation}
V_J(|w|)=\sum_mV_mJ_2(me^{-b|w|}/b)~~.
\label{30}
\end{equation}
To determine the needed coefficients $V_m$, we apply $\int_0^{\infty} d|w|
e^{-2b|w|}J_2(me^{-b|w|}/b)$ to Eq. (\ref{30}) and use the orthogonality
relations that the asymptotically well-behaved $J_2(me^{-b|w|}/b)$ modes
obey. Specifically, with the right-hand side of Eq. (\ref{16})
vanishing for these modes, the modes will then obey
\begin{equation}
\int_0^{\infty} d|w|
e^{-2b|w|}J_2(me^{-b|w|}/b)J_2(m^{\prime}e^{-b|w|}/b) =0
\label{31}
\end{equation}
when $m$ is not equal to $m^{\prime}$, with use of some standard
properties of Bessel functions obliging them to obey 
\begin{eqnarray}
\int_0^{\infty} d|w| e^{-2b|w|}J_2^2(me^{-b|w|}/b)
&=&b\int_{0}^{1/b} dx xJ_2^2(mx)
\nonumber \\
&=&b{x^2 \over 2}\left[J_2^2(mx)-J_1(mx)J_3(mx)\right] {\bigg
|}_{0}^{1/b}={J_2^2(m/b) \over 2b}~~,
\label{32}
\end{eqnarray}
when $m$ and $m^{\prime}$ are equal and $m$ is such that $J_1(m/b)$ is
zero. Armed with Eqs. (\ref{31}) and (\ref{32}) we thus find that
$V_J(|w|)$ is to be given by 
\begin{equation}
V_J(|w|)=\sum_m {2b B_m \over J_2^2(m/b)}J_2(me^{-b|w|}/b)~~,
\label{33}
\end{equation}
where the coefficients $B_m$ are given by
\begin{eqnarray}
B_m&&=\int_0^{\infty} d|w| e^{-2b|w|}V_J(|w|)J_2(me^{-b|w|}/b)
=-b\hat{V}\int ^{\beta}_{\alpha}xdxJ_2(mx)
\nonumber \\
&&=-{b\hat{V} \over m^2}\int
^{m\beta}_{m\alpha} [2J_1(x)-xJ_0(x)] 
={b\hat{V} \over m^2}[2J_0(x)+xJ_1(x)]{\bigg |}^{m\beta}_{m\alpha}
\nonumber \\
&&={b\hat{V} \over m^2}[2J_0(m\beta)+m\beta J_1(m\beta)]-
{b \hat{V}\over m^2}[2J_0(m\alpha)+m\beta J_1(m\alpha)]~~.
\label{34}
\end{eqnarray}
With every quantity which appears in Eq. (\ref{33}) now being known,
$V_J(|w|)$ can readily be plotted, and we display it in Fig. (1) as
evaluated \cite{footnote2} through the use of the first 1000
modes in the sum \cite{footnote3}. As we see, the basis is indeed capable
of generating the square step to very high accuracy, with its completeness
thus being confirmed. 

\begin{figure}
\centerline{\epsfig{file=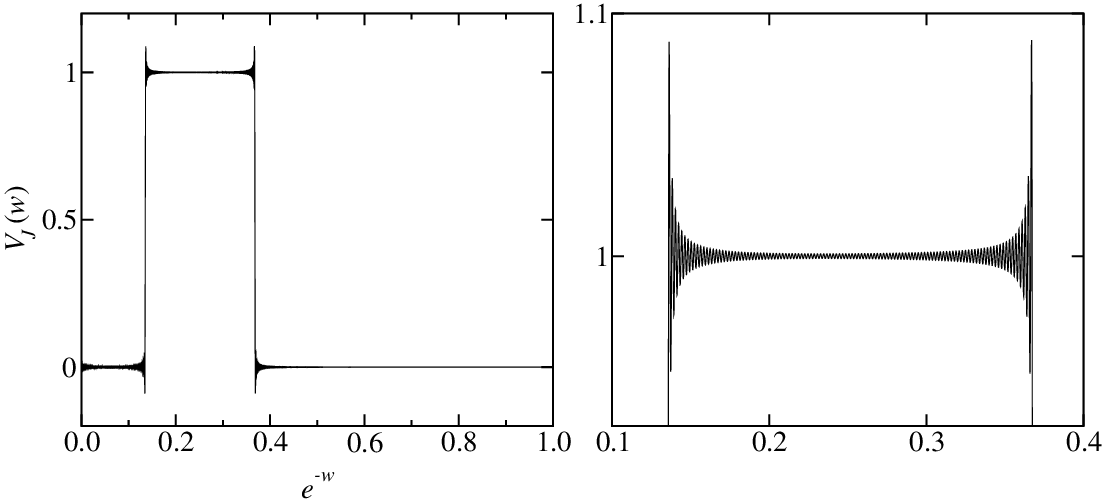,height=3.5in}}
\medskip
\caption{The left panel shows a reconstruction of the square step
$V_J(|w|)=1$, $1<|w|<2$, $V_J=0$ otherwise via the $M_4^-$ discrete
$J_2(j_ie^{-b|w|})$ mode basis, with the parameter $b$ being set equal to
one. The right panel shows a blow-up of the region near the top of the
step.}
\label{Fig. (1)}
\end{figure}

With regard to the plot in Fig. (1), as can be seen from the blow-up of
the region near the top of the step, the mode sum expressly displays the
Gibbs phenomenon associated with trying to fit a discontinuity with a
complete basis, with there being an overshoot (to near $V_J=1.1$ in the
figure) at the top of the discontinuity and an accompanying undershoot at
the bottom, an overshoot and undershoot which as required of the Gibbs
phenomenon were explicitly found to get narrower (in $|w|$) as the number
of modes in the sum was increased, but not to shorten in height, always
reaching close to $V_J=1.1$ in the figure. We regard the recovering of the
Gibbs phenomenon as a very good indicator of the reliability of our
construction, and together with the quality of the overall fit itself, as
providing very good evidence for completeness of the convergent $M_4^-$
mode basis.

\section{Completeness test for divergent $M_4^-$ modes}

To test for completeness of the divergent $Y_2(y_ie^{-b|w|})$ plus
$e^{2b|w|}$ mode basis we try to reconstruct the square step via the
expansion
\begin{equation}
V_Y(|w|)=\sum_nV_nY_2(ne^{-b|w|}/b)+V_0e^{2b|w|}~~.
\label{35}
\end{equation}
(In Eq. (\ref{35}) we  use $n$ to denote the $y_i$ zeroes of
$Y_1(y)$, and shall use $m$ to denote the $j_i$ zeroes of
$J_1(y)$.) Now while such a reconstruction might at first be thought
unlikely to succeed since every term on the right-hand side of Eq.
(\ref{35}) diverges badly in the large $|w|$ region where we need the
summation to vanish, the various terms in Eq. (\ref{35}) are not
diverging arbitrarily but, as can be seen from Eq. (\ref{28}), are
actually all diverging in exactly the same $e^{2b|w|}$ manner. In
consequence of this, we are therefore able to adjust the various
coefficients in Eq. (\ref{35}) so as to expressly cancel out the divergent
part. However, in order to get $V_Y(|w|)$ to actually vanish rather than
merely not diverge outside the step, we will also need to cancel the
finite part there as well. Thus, with each $Y_2(y)$ having a leading
behavior of the form $-4/\pi y^2-1/\pi$ at small argument, i.e. with Eq.
(\ref{35}) behaving as 
\begin{equation}
V_Y(|w|) \rightarrow e^{2b|w|}\left[V_0-{4b^2 \over \pi}\sum_n {V_n \over
n^2}\right]-{1 \over \pi}\sum_n V_n
\label{36}
\end{equation}
at large $|w|$, we need to impose the two conditions
\begin{equation}
{4b^2 \over \pi}\sum_n {V_n \over n^2}=V_0~~,~~\sum_n
V_n=0~~.
\label{37}
\end{equation}
on the coefficients, with the two leading large $|w|$ terms then being
cancelled.

Having thus taken care of the leading behavior at large $|w|$, we now try
to proceed as with our analysis of the expansion of $V_J(|w|)$ in
convergent modes. However, we cannot simply apply $\int_0^{\infty} d|w|
e^{-2b|w|}Y_2(ne^{-b|w|}/b)$ to Eq. (\ref{35}) as every overlap integral
would diverge. However, we have found it very convenient to apply
$\int_0^{\infty} d|w| e^{-2b|w|}J_2(me^{-b|w|}/b)$ to Eq. (\ref{35})
instead, where we take the $m/b$ to be the $j_i$ zeroes of $J_1(y)$. With
none of the $J_1(m/b)$ zeroes coinciding with any of the zeroes of
$Y_1(n/b)$ \cite{footnote4}, the needed overlap integrals are given (on
setting
$x=e^{-b|w|}/b$) by
\begin{eqnarray}
&&\int_0^{\infty} d|w| e^{-2b|w|}J_2(me^{-b|w|}/b)Y_2(ne^{-b|w|}/b)
=b\int_{0}^{1/b} dx xJ_2(mx)Y_2(nx)
\nonumber \\
&&~=bx\left[{nY_1(nx)J_2(mx)-mJ_1(mx)Y_2(nx) \over (m^2-n^2)}\right]
{\bigg |}_{0}^{1/b}
={ 2bm^2 \over \pi n^2(n^2-m^2)}~~,
\label{38}
\end{eqnarray}
and
\begin{eqnarray}
&&\int_0^{\infty} d|w| e^{-2b|w|}J_2(me^{-b|w|}/b)e^{2b|w|}
={1 \over b}\int_{0}^{1/b} {dx \over x}J_2(mx)
\nonumber \\
&&~=-{1 \over b}\int_{0}^{1/b} dx{d \over
dx}\left({J_1(mx)
\over mx}\right) ={1\over 2b}~~,
\label{39}
\end{eqnarray}
overlap integrals which despite the badly divergent behavior of the
$Y_2(y)$ and $e^{2b|w|}$ are nonetheless actually finite due to the
compensating convergent behavior of the $J_2(y)$. On thus applying
$\int_0^{\infty} d|w| e^{-2b|w|}J_2(me^{-b|w|}/b)$ to Eq.
(\ref{35}), we find that for the square step $V_Y(|w|)=\hat{V}$,
$\alpha \leq e^{-b|w|}/b \leq \beta$, $V_Y(|w|)=0$ otherwise, the
expansion coefficients must thus obey
\begin{eqnarray}
{V_0 \over 2b}+{2b \over \pi}\sum _n V_n{m^2 \over 
n^2(n^2-m^2)}&&={V_0 \over 2b}+{2b \over \pi}\sum _n V_n\left[ {1
\over  (n^2-m^2)}-{1 \over n^2}\right]
\nonumber \\
&&={2b \over \pi}\sum _n {V_n
\over  (n^2-m^2)}=B_m
~~
\label{40}
\end{eqnarray}
for all $m$, where the $B_m$ are given by  
\begin{eqnarray}
B_m&&=-b\hat{V}\int_{\alpha}^{\beta} dx x J_2(mx)
={b\hat{V} \over
m^2}[2J_0(mx)+mxJ_1(mx)]{\bigg |}_{\alpha}^{\beta} 
\nonumber \\
&&={b\hat{V} \over
m^2}[2J_0(m\beta)+m\beta J_1(m\beta)]-{b\hat{V} \over
m^2}[2J_0(m\alpha)+m\alpha J_1(m\alpha)]
~~.
\label{41}
\end{eqnarray}
With the $B_m$ being given in closed form, Eq. (\ref{40}) is thus a
set of $N$ equations for $N$ unknowns and can be viewed as an eigenvalue
equation for the $V_n$. (While the $J_2(me^{-b|w|}/b)Y_2(ne^{-b|w|}/b)$
overlap integrals of Eq. (\ref{38}) are finite, the
$J_2(me^{-b|w|}/b)$ and $Y_2(ne^{-b|w|}/b)$ modes are not orthogonal, with
Eq. (\ref{40}), unlike Eq. (\ref{33}), thus not being diagonal in its
indices.) The
$V_n$ coefficients can thus be determined, and on being found
to be finite and rapidly oscillating in sign, lead, for the case of the
first 1000 modes in the basis, to the plot displayed in Fig. (2) (i.e. we
restrict to the first 1000
$y_i$ and the first 1000 $j_i$ in Eq. (\ref{40})). As Fig. (2) thus 
indicates, and quite spectacularly so, the divergent mode basis is every
bit as capable of reconstructing the square step as the convergent one and
every bit as capable of recovering the Gibbs phenomenon, and is thus
every bit as complete \cite{footnote5}. It is thus invalid to use
normalizability as a criterion for discarding modes as non-normalizable
modes are fully capable of serving as a complete basis for constructing
localized packets \cite{footnote6}. As a final comment, we recall that for
the harmonic oscillator wave equation there are two sets of solutions,
the sines and the cosines, and both sets are complete. It is hence
perfectly reasonable to expect other second order wave equations to
also have two complete sets of bases even if one of them consists entirely
of divergent modes.  

\begin{figure}
\centerline{\epsfig{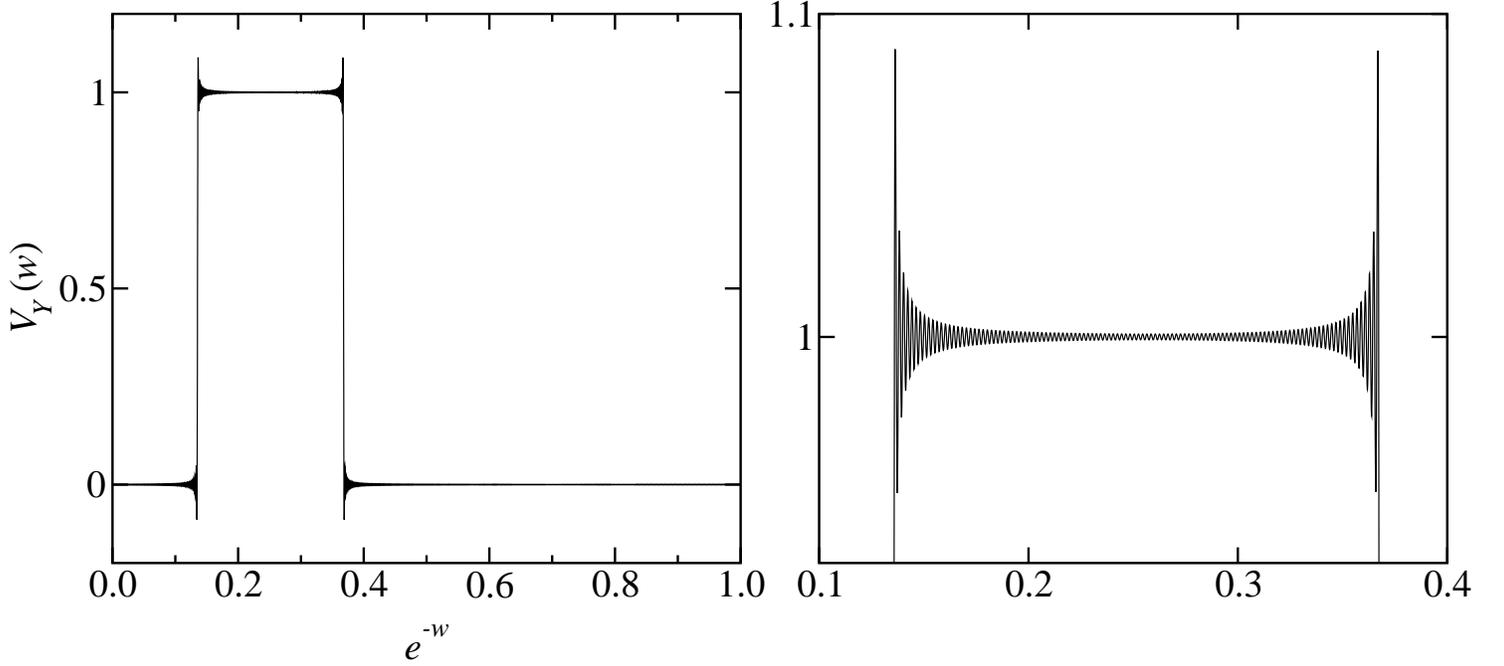}}
\medskip
\caption{The left panel shows a reconstruction of the square step
$V_Y(|w|)=1$, $1<|w|<2$,
$V_Y =0$ otherwise via the $M_4^-$ discrete $Y_2(y_ie^{-b|w|})$ plus
$e^{2b|w|}$ mode basis, with the parameter $b$ being set equal to
one. The right panel shows a blow-up of the region near the top of the
step.}
\label{Fig. (2)}
\end{figure}

\section{Completeness tests for the anti-de Sitter brane cases}

\subsection{The basis modes}

For $AdS_4^+$ brane world with warp factor $e^{A(|w|)}=H{\rm
cosh}(\sigma-b|w|)/b$ where ${\rm
cosh}\sigma=b/H$, the transformation $y={\rm
tanh}(b|w|-\sigma)$ brings Eq. (\ref{13}) to the form 
\begin{equation}
\left[(1-y^2){d^2 \over dy^2}-2y{d \over dy}+\nu(\nu+1)-{4 \over
(1-y^2)}\right]f_m(y)=0~~.
\label{42}
\end{equation}
where we have introduced the convenient parameter $\nu$ defined by
\begin{equation}
\nu=\left({9\over 4}+{m^2\over H^2}\right)^{1/2}-{1\over 2}~~,~~
{m^2 \over H^2}=(\nu-1)(\nu+2)~~.
\label{43}
\end{equation}
Equation (\ref{42}) is recognized as an associated Legendre equation, with
its solutions being the associate Legendre functions of the first and
second kind, so that for $m \neq 0 $ (viz. $\nu \neq 1$) we can set 
\begin{equation}
f_m(y)=\alpha_m P^2_{\nu}(y)+\beta_m Q^2_{\nu}(y)~~.
\label{44}
\end{equation}
This solution also applies to one of the $m=0$ solutions as well, viz.
$Q_1^2(y)$, a quantity which can be written in terms of the warp factor as
$Q_1^2(y)=2/(1-y^2)=2{\rm cosh}^2(b|w| -\sigma)=2b^2e^{2A(|w|)}/H^2$, but
misses one other solution since $P_1^2(y)$ is kinematically zero. This
second
$m=0$ solution can be found by setting $\nu=1$ in Eq. (\ref{42}) and
solving it directly, to yield
\begin{equation}
f_{0}(y)=\alpha_{0}\bigg{(}\frac{2}{(1+y)}-y\bigg{)}+
\beta_0Q^2_{1}(y)~~.
\label{45}
\end{equation}
Requiring the modes to also obey the junction condition of Eq.
(\ref{14}) then restricts them according to
\begin{equation}
\alpha_m P^1_{\nu}(-{\rm tanh}\sigma)+\beta_m 
Q^1_{\nu}(-{\rm tanh}\sigma)=0~~,~~\alpha_0=0~~,
\label{46}
\end{equation}
to thus define the $AdS_4^+$ brane world basis modes.

As functions, all of the $P^1_{\nu}(y)$, $P^2_{\nu}(y)$, $Q^1_{\nu}(y)$
and $Q^2_{\nu}(y)$ possess a cut in the complex $y$ plane which can be
located to run from $y=-\infty$ to $y=1$. For the $AdS_4^+$ brane world
the parameter $y={\rm tanh}(b|w|-\sigma)$ lies in the range $-{\rm
tanh}\sigma \leq y \leq 1$, and so in this range the 
Legendre functions have to to be evaluated on the cut (as the real
$P_{\nu}^{\mu}(y)=(1/2)[e^{i\pi\mu/2}P_{\nu}^{\mu}(y+i\epsilon)
+e^{-i\pi\mu/2}P_{\nu}^{\mu}(y-i\epsilon)]$, 
$Q_{\nu}^{\mu}(y)=(e^{-i\pi\mu}/2)[e^{-i\pi\mu/2}Q_{\nu}^{\mu}(y+i\epsilon)
+e^{i\pi\mu/2}Q_{\nu}^{\mu}(y-i\epsilon)]$) where they can then be power
series expandable via their relation to hypergeometric functions to yield
\begin{eqnarray}
P_{\nu}^m(y)&=&{(-1)^{m}\Gamma(\nu +m +1) \over
2^m m!\Gamma(\nu -m+1)}(1-y^2)^{m/2}F(\nu +m+1,-\nu +m;m+1;(1-y)/2)
\nonumber \\
&=&
{(-1)^{m}\Gamma(\nu +m +1) \over
2^m m!\Gamma(\nu -m+1)}(1-y^2)^{m/2}\bigg{[}1
+{(\nu+m+1)(-\nu+m) \over (m+1)1!}{(1-y) \over 2}
\nonumber \\
&~&
+{(\nu+m+1)(\nu+m+2)(-\nu+m)(-\nu+m+1) \over (m+1)(m+2)2!}{(1-y)^2 \over
2^2}+...\bigg{]}~~,
\nonumber \\
Q_{\nu}^m(y)&=&{e^{im\pi}2^{\nu}\Gamma(\nu +1)\Gamma(\nu +m +1) \over
\Gamma(2\nu +2)(1+y)^{\nu +1-m/2}(1-y)^{m/2}}F(\nu -m+1,\nu +1;2\nu
+2;2/(1+y))
\nonumber \\
&=&{e^{im\pi}2^{\nu}\Gamma(\nu +1)\Gamma(\nu +m +1) \over
(1+y)^{\nu +1-m/2}(1-y)^{m/2}}\bigg[
{\Gamma(m) \over \Gamma(\nu
+1)\Gamma(\nu+m+1)}
\nonumber \\
&~&\times \sum_{n=0}^{m-1}{(\nu-m+1)_n(\nu+1)_n
\over
(1-m)_nn!}{(y-1)^n\over (y+1)^n}
\nonumber \\
&~&+{(-1)^m(y-1)^m \over \Gamma(\nu
-m +1)\Gamma(\nu+1)(y+1)^m}\sum_{n=0}^{\infty}{(\nu +1)_n(\nu +m+1)_n
\over (n+m)!n!}{(y-1)^n \over (y+1)^n}
\nonumber \\
&&\times \bigg{[}\psi(n+1)+\psi(n+m+1) -\psi(\nu +1 +n) -\psi(\nu +m +1
+n) -{\rm log}\left(1-y\over 1+y\right)\bigg{]}\bigg{]}
\nonumber \\
\label{47} 
\end{eqnarray}
when $\mu$ is a general positive integer $m$. (In Eq. (\ref{47})
$\psi(y)$ denotes $ (d\Gamma(y)/dy)/\Gamma(y)$ and
$(a)_n$ denotes $\Gamma(a+n)/\Gamma(a)$). From Eq. (\ref{47}) we see that
in the
$-{\rm tanh}\sigma \leq y \leq 1$ range of interest the
$P^2_{\nu}(y)$ are well-behaved, behaving as
$y$ approaches one from below (viz. as $|w|\rightarrow \infty$) as   
\begin{equation}
P^2_{\nu}(y\rightarrow 1) \rightarrow P(\nu)\left[(1-y)
-{(1-y)^2(\nu^2+\nu-3)\over 6}\right]
\label{48}
\end{equation}
where 
\begin{equation}
P(\nu)={\nu(\nu^2-1)(\nu+2)\over 4}~~,
\label{49}
\end{equation}
to thus be fully normalizable and have finite normalization 
\begin{equation}
N_{\nu}=\int_{-\infty}^{\infty}dwe^{-2A}[P_{\nu}^2(|w|)]^2=
2\int_{0}^{\infty}d|w|e^{-2A}[P_{\nu}^2(|w|)]^2={2b \over H^2}\int
_{-{\rm tanh}\sigma}^1 dy [P_{\nu}^2(y)]^2~~.
\label{50}
\end{equation}
However, unlike the $P^2_{\nu}(y)$, the $Q_{\nu}^2(y)$ all found to
diverge at $y=1$, behaving there as 
\begin{equation}
Q^2_{\nu}(y\rightarrow 1) \rightarrow 
{1\over (1-y)}+{(\nu^2+\nu-1)\over 2}+O\left((1-y){\rm
ln}(1-y)\right)~~,
\label{51}
\end{equation}
and thus in the $AdS_4^+$ brane world none of the $Q_{\nu}^2(y)$, and
particularly the massless $Q_{1}^2(y)$ graviton, are
normalizable. We shall thus seek to construct complete bases in both the
normalizable and non-normalizable sectors.

\subsection{Completeness test for convergent $AdS_4^+$ modes}

To construct a complete basis out of normalizable modes alone requires
that the normalizable $P_{\nu}^2(y)$ satisfy Eq. (\ref{46}) all on their
own, with the eigenmodes then needing to satisfy 
\begin{equation}
P^{1}_{\nu}(-{\rm tanh}\sigma)=P^{1}_{\nu}(-(1-H^2/b^2)^{1/2})=0~~.
\label{52}
\end{equation}
For arbitrary $\sigma$ the solutions to Eq. (\ref{52}) cannot be
written down in a closed form, but on noting that for one particular
value of $\sigma$, viz. $\sigma=0$ (i.e. $H=b$), $P^{1}_{\nu}(0)$ is known
in closed form as
\begin{equation}
P^{1}_{\nu}(0)=\frac{2\pi^{1/2}}{\Gamma(\nu/2+1/2)
\Gamma(-\nu/2)}~~,
\label{53}
\end{equation}
to thus be zero at $\nu=2,4,6,...$, we see that on solving for an
arbitrary given $\sigma$ numerically an infinite discrete set of allowed
$\nu$ values will then be found to ensue \cite{footnote7}. The
normalizable mode sector of $AdS_4^+$ is thus discrete and infinite, a
result first obtained in \cite{Karch2001} by directly numerically solving
Eq. (\ref{13}).

To test for completeness of the normalizable $AdS_4^+$ mode basis we need
to  find a set of coefficients $V_m$ for which the expansion
\begin{equation}
V_P= \sum_m V_{m}P_{\nu}^2(y)
\label{54}
\end{equation}
reproduces the square step $V_P=\hat{V}$ when $|w_1|<|w|<|w_2|$, $V_P=0$
otherwise. With the $P_{\nu}^2(y)$ modes being orthogonal, the
coefficients are readily given as $V_m=B_m/N_{\nu}$ where $N_{\nu}$ is
the normalization factor given in Eq. (\ref{50}), where $m$ and $\nu$ are
related as in Eq. (\ref{43}), and where some standard properties of the
associated Legendre functions allow
$B_m$ to be written as 
\begin{eqnarray}
B_{m}&=&\hat{V}\int_{|w_1|}^{|w_2|}d|w|e^{-2A}P_{\nu}^2(|w|)=
\frac{\hat{V}b}{H^2}\int_{y_1}^{y_2}dyP^2_{\nu}(y)
=\frac{\hat{V}b}{H^2}\int_{y_1}^{y_2}dy(1-y^2)\frac{d^2P_{\nu}(y)}{dy^2}
\nonumber \\
&=&\frac{\hat{V}b}{H^2}\int_{y_1}^{y_2}dy\left[\frac{d}{dy}\left[(2-\nu)y
P_{\nu}+\nu P_{\nu -1}\right]-2P_{\nu}\right]
\nonumber \\
&=&\frac{\hat{V}b}{H^2}\int_{y_1}^{y_2}dy\frac{d}{dy}\left[
{(2-\nu)\over (2\nu +1)}
[(\nu  +1)P_{\nu +1}+\nu P_{\nu -1}]+\nu P_{\nu -1}
-{2 \over (2\nu +1)}
\left(P_{\nu +1}-P_{\nu -1}\right)\right]
\nonumber \\
&=&\frac{\hat{V}b}{H^2}\left[\frac{(\nu +1)(\nu +2)P_{\nu -1}-
\nu(\nu -1)P_{\nu +1}}{2\nu +1}\right]{\bigg |}_{y_1}^{y_2}~~.
\label{55} 
\end{eqnarray}
With every quantity which appears in Eq. (\ref{54}) now being known,
$V_P(|w|)$ can readily be plotted, and we display it in Fig. (3) as
evaluated through the use of the first 1000 modes in the
sum. As we see, the basis is indeed capable of generating
the square step to very high accuracy, and with it expressly
displaying the Gibbs phenomenon \cite{footnote8}, its completeness is
thus  confirmed. 

\begin{figure}
\centerline{\epsfig{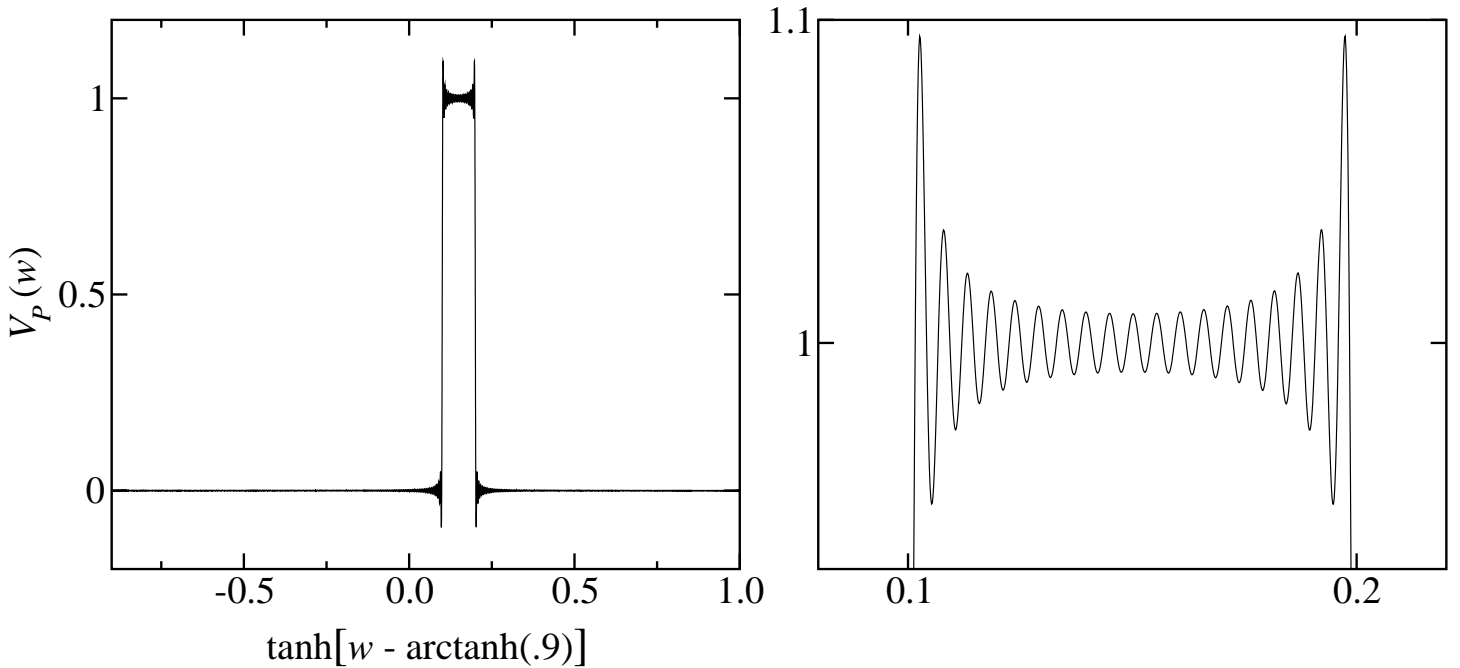}}
\medskip
\caption{The left panel shows a reconstruction of the square step
$V_P(w)=1$,  $0.1 \leq {\rm
tanh} (b|w|-{\rm arctanh}(0.9)) \leq 0.2$, $V_P(w)=0$ otherwise, via the
$AdS_4^+$ discrete $P_{\nu}^2({\rm tanh} (b|w|-\sigma))$ mode basis in the
typical case where ${\rm tanh}\sigma =0.9$, $H/b=0.436$, and $b=1$. The
right panel shows a blow-up of the region near the top of the step.}
\label{Fig. (3)}
\end{figure}

\subsection{Completeness test for divergent $AdS_4^+$ modes}

With the massless $AdS_4^+$ graviton with divergent warp factor wave
function $f_0(y)=\beta_0Q_1^2(y)=2\beta_0/(1-y^2)$ obeying the junction
condition, it could also belong to a complete basis of divergent
$Q_{\nu}^2(y)$ modes (modes which according to Eq. (\ref{51}) actually
diverge in precisely the same $1/(1-y)$ way near $y=1$ as the massless
graviton itself) if the $Q_{\nu}^2(y)$ modes were to satisfy the junction
condition on their own, i.e. if they were to obey
\begin{equation}
Q^{1}_{\nu}(-{\rm tanh}\sigma)=Q^{1}_{\nu}(-(1-H^2/b^2)^{1/2})=0~~. 
\label{56}
\end{equation}
With Eq. (\ref{56}) being found to possess an infinite set of discrete
solutions for the arbitrary $\sigma$ \cite{footnote9}, we shall thus seek
to expand the localized square step $V_Q=\hat{V}$ when $|w_1|\leq
|w|\leq |w_2|$, $V_Q=0$ otherwise, in terms of these solutions as 
\begin{equation}
V_Q= \sum_n V_{n}Q_{\nu}^2(y)+\frac{V_0}{1-y^2}~~.
\label{57}
\end{equation}
(For clarity we use $n^2$ here to denote the squared masses
$n^2/H^2=(\nu-1)(\nu+2)$ of the $Q_{\nu}^2(y)$ sector modes, and use
$m^2$ for the $P_{\nu}^2(y)$ sector.) Given the asymptotic limit
exhibited in Eq. (\ref{51}), in order to first cancel both the leading
$1/(1-y)$ term and the next to leading $O(1)$ term from the right-hand
side of Eq. (\ref{57}), we must constrain the $V_n$ coefficients
according to 
\begin{equation}
\sum_n V_{n}+\frac{V_0}{2}=0~~,~~\frac{1}{2}\sum_n
V_n(\nu^2+\nu-1)+\frac{V_0}{4}=0~~,
\label{58}
\end{equation}
to thus enable us to reexpress the square step
expansion as 
\begin{equation}
V_Q= \sum_n V_{n}\left[Q_{\nu}^2(y)-\frac{2}{1-y^2}\right]~~,
\label{59}
\end{equation}
as subject to the constraint
\begin{equation}
\sum_n V_{n}\left[\nu^2+\nu-2\right]=\sum_n
V_{n}\frac{n^2}{H^2}=0~~.
\label{60}
\end{equation}

While we cannot apply $\int_0^{\infty} d|w| e^{-2A}Q_{\nu}^2(y)$ to Eq.
(\ref{59}) as every overlap integral would diverge, finite overlap
integrals are obtained if we instead apply $\int_0^{\infty} d|w|
e^{-2A}P_{\nu^{\prime}}^2(y)$, where we use $\nu^{\prime}$ to label
the $P_{\nu^{\prime}}^2(y)$ sector so that its squared masses are given
by $m^2/H^2=(\nu^{\prime}-1)(\nu^{\prime}+2)$. With none of the
$P^{1}_{\nu^{\prime}}(-{\rm tanh}\sigma)$ and $Q^{1}_{\nu}(-{\rm
tanh}\sigma)$ zeroes being found to coincide, via. Eqs. (\ref{16}),
(\ref{48}) and (\ref{51}) the needed overlap integrals are found to be of
the form
\begin{equation}
\int_0^{\infty}d|w|e^{-2A}P_{\nu^{\prime}}^2(y)Q_{\nu}^2(y)
=\frac{4bP(\nu^{\prime})}{(m^2-n^2)}
\label{61}
\end{equation}
($P(\nu^{\prime})$ is given in Eq. (\ref{49})), and are indeed finite,
just as required. With the overlap integral which involves the massless
graviton mode being given by
\begin{equation}
\int_0^{\infty}d|w|e^{-2A}\frac{P_{\nu^{\prime}}^2(y)}{(1-y^2)}
=\frac{2bP(\nu^{\prime})}{m^2}~~,
\label{62}
\end{equation}
the application of $\int_0^{\infty} d|w| e^{-2A}P_{\nu^{\prime}}^2(y)$ to
Eq. (\ref{59}) thus yields 
\begin{equation}
4b\sum_nV_nP(\nu^{\prime})
\left[\frac{1}{(m^2-n^2)} -\frac{1}{m^2}\right]
=
\sum_nV_n 
{ b(m^2+2H^2)n^2\over H^4(m^2-n^2)}=B_m~~,
\label{63}
\end{equation}
where $B_m$ is the same function that was already given earlier in Eq.
(\ref{55}). 

\begin{figure}
\centerline{\epsfig{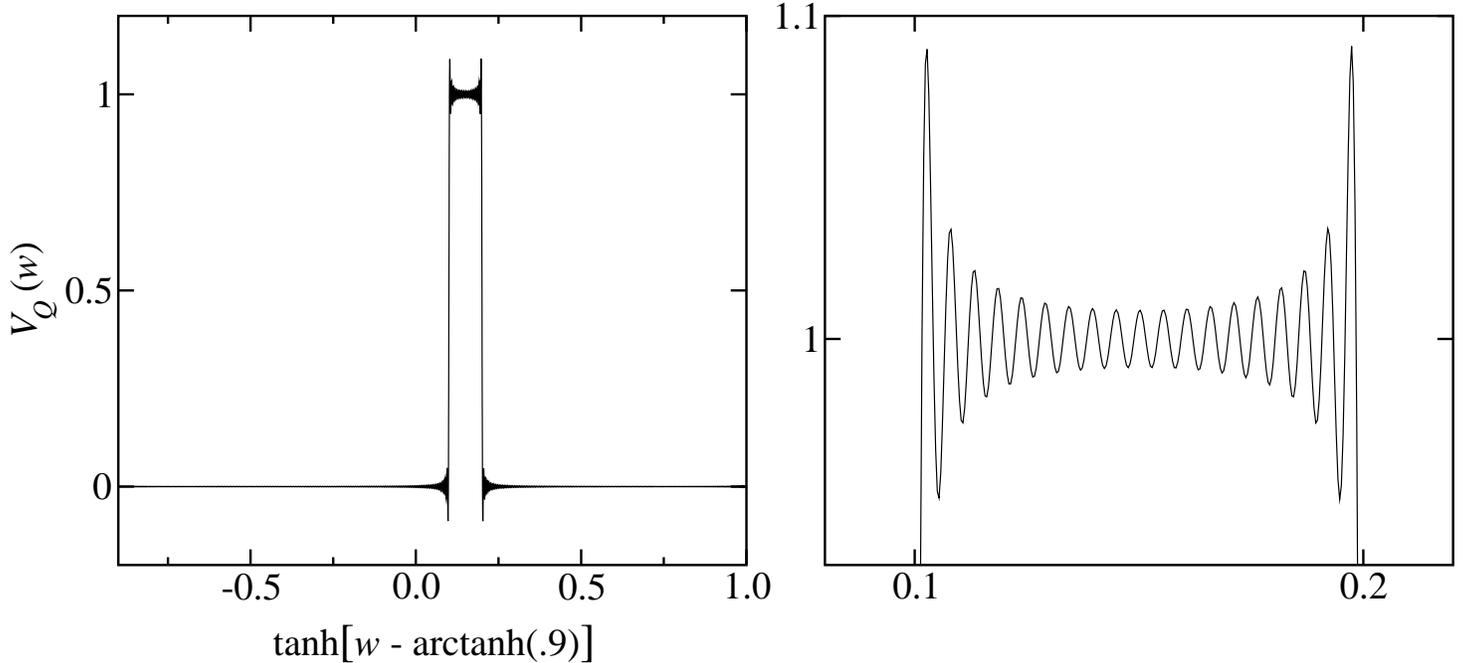}}
\medskip
\caption{The left panel shows a reconstruction of the square step
 $V_Q(w)=1$,  $0.1 \leq {\rm
tanh} (b|w|-{\rm arctanh}(0.9)) \leq 0.2$, $V_Q(w)=0$ otherwise, via the
$AdS_4^+$ discrete $Q_{\nu}^2({\rm tanh} (b|w|-\sigma))$ plus ${\rm
cosh}^2({\rm tanh} (b|w|-\sigma))$ mode basis in the typical case where
${\rm tanh}\sigma =0.9$,
$H/b=0.436$, and
$b=1$. The right panel shows a blow-up of the region near the top of the
step.}
\label{Fig. (4)}
\end{figure}

Given Eq. (\ref{63}), the $V_n$ coefficients can now be found numerically,
and lead, for the case of the first 1000 modes in the basis to the plot
displayed in Fig. (4) (i.e. we restrict to the first 1000
$P^{1}_{\nu^{\prime}}(-{\rm tanh}\sigma)$ zeroes and the first 1000
$Q^{1}_{\nu}(-{\rm tanh}\sigma)$ zeroes). As Fig. (4) thus indicates, the
divergent mode basis is every bit as capable of reconstructing the square
step as the convergent one and every bit as capable of recovering the
Gibbs phenomenon, and is thus every bit as complete \cite{footnote10}.
Once again then we see that it is invalid to use normalizability as a
criterion for discarding modes, and in this regard we differ from the
view of \cite{Karch2001} that it is permissible to discard modes such as
the massless $AdS_4^+$ graviton simply because they are not
normalizable \cite{footnote11}.

\section{Completeness tests for the de Sitter brane cases}

\subsection{The basis modes}

For $dS_4^{\pm}$ brane worlds with warp factor $e^{A(|w|)}=H{\rm
sinh}(\sigma\mp b|w|)/b$ where ${\rm sinh}\sigma=b/H$, the transformation
$y={\rm coth}(\sigma \mp b|w|)$ brings Eq. (\ref{13}) to the form 
\begin{equation}
\left[(1-y^2){d^2 \over dy^2}-2y{d \over dy}+\nu(\nu+1)-{4 \over
(1-y^2)}\right]f_m(y)=0~~.
\label{64}
\end{equation}
where we have introduced the convenient parameter $\nu$ defined by
\begin{equation}
\nu=\left({9\over 4}-{m^2\over H^2}\right)^{1/2}-{1\over 2}~~,~~
{m^2 \over H^2}=(1-\nu)(\nu+2)~~.
\label{65}
\end{equation}
Recognizing Eq. (\ref{64}) to be the previously discussed associated
Legendre equation, its $m \neq 0 $ (viz. $\nu \neq 1$) solutions are
given as
\begin{equation}
f_m(y)=\alpha_m P^2_{\nu}(y)+\beta_m Q^2_{\nu}(y)~~.
\label{66}
\end{equation}
while its $\nu=1$ solutions are of the form
\begin{equation}
f_{0}(y)=\alpha_{0}\bigg{(}\frac{2}{(1+y)}-y\bigg{)}+
\beta_0Q^2_{1}(y)~~.
\label{67}
\end{equation}
Requiring the modes to also obey the junction condition of Eq.
(\ref{14}) then restricts them according to
\begin{equation}
\alpha_m P^1_{\nu}({\rm coth}\sigma)+\beta_m 
Q^1_{\nu}({\rm coth}\sigma)=0~~,~~\alpha_0=0~~,
\label{68}
\end{equation}
to thus define the $dS_4^{\pm}$ brane-world basis modes.

While the $dS_4^+$ and $dS_4^-$ basis modes are quite similar to each
other in their generic structure, they differ from each other
significantly in one crucial regard. Specifically, unlike the $dS_4^-$
warp factor $e^{A(|w|)}=H{\rm sinh}(\sigma+ b|w|)/b$ which never vanishes
($\sigma$ having been defined to be positive), the $dS_4^+$ warp
factor $e^{A(|w|)}=H{\rm sinh}(\sigma- b|w|)/b$ has a zero at
$b|w|=\sigma$. With a null signal taking an infinite amount of
time to travel from the brane to the location of this zero, this zero
serves as a horizon for an observer on the brane \cite{Garriga2000}, with
the brane observer only being sensitive to fluctuation modes in the
$\sigma \geq b|w| \geq 0$ region. With the  $dS_4^+$ parameter $y={\rm
coth}(\sigma - b|w|)$ lying in the range ${\rm coth \sigma} \leq y \leq
\infty$, we see that
$y$ is infinite at the $dS_4^+$ horizon. Then, with the associated
Legendre functions behaving as
$P^2_{\nu}(y) \rightarrow O(y^{\nu})+O(y^{-\nu-1})$, $Q^2_{\nu}(y)
\rightarrow O(y^{-\nu-1})$ as $y \rightarrow \infty$, the 
$\nu=1$ massless $dS_4^+$ graviton and all $dS_4^+$ modes with complex
$\nu=-1/2\pm i(m^2/H^2-9/4)^{1/2}$ will be normalizable within the horizon
\cite{footnote12}. With the massless graviton and a massive continuum of
modes with $m^2/H^2 \geq 9/4$ which satisfy the junction condition of Eq.
(\ref{68}) by an interplay (of the real $P_{\nu}^{2}(y)$ and
the real part of $Q_{\nu}^{2}(y)$) thus providing a conventional
continuum normalized complete basis in the sense of Eqs. (\ref{1}) to
(\ref{5}), as with the $M_4^+$ brane world, in the $dS_4^+$ brane world
there is no need to test explicitly for completeness.

However, for $dS_4^-$ the situation is quite different since there is now
no vanishing of the warp factor and no horizon, with the coordinate
$|w|$ now extending all the way to infinity, and with the parameter
$y={\rm coth}(\sigma +b|w|)$ instead now lying in the $1 \leq y
\leq {\rm coth \sigma}=(1+H^2/b^2)^{1/2}$ range. Unlike the previously
discussed $AdS_4^+$ brane world case where $y$ approached one from below
as
$|w|$ went to infinity, in the $dS_4^-$ case $y$ instead approaches one
from above in the large $|w|$ limit, with Eqs. (\ref{48}) and (\ref{51})
having to be replaced by the limits
\begin{eqnarray}
&&P^2_{\nu}(y\rightarrow 1) \rightarrow P(\nu)\left[(y-1)
+{(y-1)^2(\nu^2+\nu-3)\over 6}\right]
\nonumber \\
&&Q^2_{\nu}(y\rightarrow 1) \rightarrow 
{1\over (y-1)}-{(\nu^2+\nu-1)\over 2}+O\left((y-1){\rm
ln}(y-1)\right)~~,
\label{69}
\end{eqnarray}
where $P(\nu)=\nu(\nu^2-1)(\nu+2)/ 4$ is as given in Eq. (\ref{49}). Since
the $P_{\nu}^2(y)$ are well behaved at $y=1$, while the $Q_{\nu}^2(y)$
diverge there, as with the $AdS_4^+$ case, the normalizable sector will
consist of all $P_{\nu}^2({\rm coth}(\sigma +b|w|))$ modes which satisfy
the junction condition on their own according to
\begin{equation}
P^1_{\nu}({\rm coth}\sigma)=
P^1_{\nu}((1+H^2/b^2)^{1/2})=0~~,
\label{70}
\end{equation}
while the non-normalizable sector will consist of the divergent warp
factor wave function $Q_{1}^2({\rm coth}(\sigma +b|w|))$
(=$2/(y^2-1)$ in $y>1$) massless graviton and all massive $Q_{\nu}^2({\rm
coth}(\sigma +b|w|))$ modes which obey
\begin{equation}
Q^1_{\nu}({\rm coth}\sigma)=
Q^1_{\nu}((1+H^2/b^2)^{1/2})=0~~. 
\label{71}
\end{equation}
While this pattern is thus quite similar to the situation found in the
$AdS_4^+$ case, the $dS_4^-$ brane world differs from it in one key
regard, namely that the parameter $y$ is required to be greater or equal
to one rather than less than or equal to it, and thus the completeness of
its mode bases requires independent testing.

\subsection{Completeness test for convergent $dS_4^-$ modes}

With the general Eq. (\ref{16}) taking the form 
\begin{eqnarray}
&&\left({m_1^2 \over H^2}-{m_2^2
\over H^2}\right)\int_1^{{\rm coth}\sigma}dyf_{m_1}(y)f_{m_2}(y)
\nonumber \\
&&~=
\lim_{y \rightarrow 1}
\left[(y^2-1)f_{m_2}(y){df_{m_1}(y) \over
dy}-(y^2-1)f_{m_1}(y){df_{m_2}(y)
\over dy}\right]
\label{72}
\end{eqnarray}
in the $dS_4^-$ case, and with the $P_{\nu}^2(y)$ modes
behaving near $y=1$ as in Eq. (\ref{69}), the $P_{\nu}^2(y)$ modes
form an orthonormal basis, and we can normalize thus them according to 
\begin{equation}
N_{\nu}=\int_{-\infty}^{\infty}dwe^{-2A}[P_{\nu}^2(|w|)]^2={2b \over
H^2}\int _{1}^{{\rm coth}\sigma} dy [P_{\nu}^2(y)]^2~~.
\label{73}
\end{equation}
With the ${\rm coth} \sigma$ argument of $P^1_{\nu}({\rm coth}\sigma)$ in
Eq. (\ref{70}) being greater than one, the $P^1_{\nu}({\rm coth}\sigma)=0$
condition has no solutions with real $\nu$. Rather, all of its solutions
are of the form $\nu=-1/2+i\lambda$ where $\lambda$ is real
and discrete \cite{footnote13}. According to Eq. (\ref{65}), for such
solutions the associated squared masses obey $m^2/H^2=9/4+\lambda^2$ and
are thus nicely positive. Additionally, as noted previously, for the
particular choice of $\nu=-1/2+i\lambda$, the $P_{\nu}^{2}(y)$ mode wave
functions themselves are real. 

Having now explicitly identified the $dS_4^-$ normalizable mode basis, to
test for completeness we need to find a set of coefficients
$V_m$ for which the expansion
\begin{equation}
\hat{V}_P= \sum_m V_{m}P_{\nu}^2(y)
\label{74}
\end{equation}
reproduces the square step $\hat{V}_P=\hat{V}$ when $|w_1|<|w|<|w_2|$,
$\hat{V}_P=0$ otherwise. With the $P_{\nu}^2(y)$ modes being orthogonal,
the coefficients are readily given as $V_m=B_m/N_{\nu}$ where $N_{\nu}$ is
the normalization factor given in Eq. (\ref{73}), where $m$ and $\nu$ are
related as in Eq. (\ref{65}), and where the $B_m$ are given as
\begin{eqnarray}
B_{m}&=&\hat{V}\int_{|w_1|}^{|w_2|}d|w|e^{-2A}P_{\nu}^2(|w|)=
-\frac{\hat{V}b}{H^2}\int_{y_1}^{y_2}dy P^2_{\nu}(y)
=-\frac{\hat{V}b}{H^2}\int_{y_1}^{y_2}dy(y^2-1)\frac{d^2P_{\nu}(y)}{dy^2}
\nonumber \\
&=&-\frac{\hat{V}b}{H^2}\int_{y_1}^{y_2}dy\left[\frac{d}{dy}\left[(\nu
-2)y P_{\nu}-\nu P_{\nu -1}\right]+2P_{\nu}\right]
\nonumber \\
&=&-\frac{\hat{V}b}{H^2}\int_{y_1}^{y_2}\frac{d}{dy}\left[
{(\nu -2)\over (2\nu +1)}
[(\nu  +1)P_{\nu +1}+\nu P_{\nu -1}]-\nu P_{\nu -1}
+{2 \over (2\nu +1)}
\left(P_{\nu +1}-P_{\nu -1}\right)\right]
\nonumber \\
&=&-\frac{\hat{V}b}{H^2}\left[\frac{
[\nu(\nu -1)P_{\nu +1}- (\nu +1)(\nu +2)P_{\nu -1}]}{2\nu +1}\right]{\bigg
|}_{y_1}^{y_2}~~.
\label{75} 
\end{eqnarray}
Given Eq. (\ref{75}),
$\hat{V}_P(|w|)$ can readily be plotted, and we display it in Fig. (5)
as evaluated through the use of the first 500 modes in the
sum. As we see, the basis is indeed capable of generating
the square step to very high accuracy, and with it also nicely
displaying the Gibbs phenomenon, its completeness is thus 
confirmed. 

\begin{figure}
\centerline{\epsfig{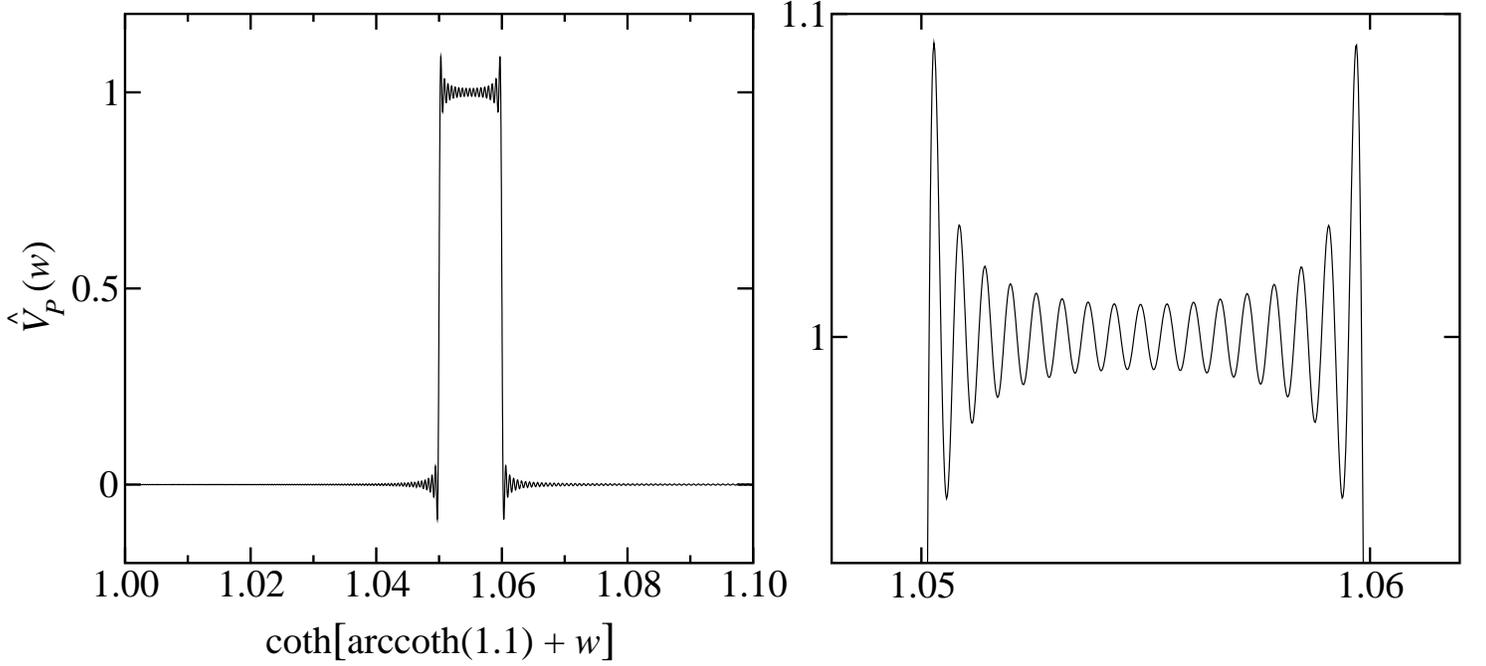}}
\medskip
\caption{The left panel shows a reconstruction of the square step
$\hat{V}_P(w)=1$,  $1.05 \leq
{\rm coth} ({\rm arccoth}(1.1) +b|w|) \leq 1.06$, $\hat{V}_P(w)=0$
otherwise, via the $dS_4^-$ discrete $P_{\nu}^2({\rm coth} (\sigma
+b|w|))$ mode basis in the typical case where ${\rm coth}\sigma =1.1$,
$H/b=0.458$, and $b=1$. The right panel shows a blow-up of the region
near the top of the step.}
\label{Fig. (5)}
\end{figure}

\subsection{Completeness test for divergent $dS_4^-$ modes}

As with the $P^{1}_{\nu}({\rm coth}\sigma)=0$ condition, the solutions
to $Q^{1}_{\nu}({\rm coth}\sigma)=0$ are also all of the form
$\nu=-1/2+i\lambda$ where $\lambda$ is again real and discrete, with the
solutions to $P^{1}_{\nu}({\rm coth}\sigma)=0$ and $Q^{1}_{\nu}({\rm
coth}\sigma)=0$ being found to interlace each other \cite{footnote14}.
With it being only the real parts of the $Q_{\nu}^2(y)$ wave functions
with $\nu=-1/2+i\lambda$ and $y$ real which are independent of the real
$P_{\nu}^2(y)$, the non-normalizable $dS_4^-$ brane world mode basis
consists of the massless graviton with its real warp factor wave function
plus the real parts of the $Q_{\nu}^2(y)$ wave functions with the
appropriate
$\nu=-1/2+i\lambda$. Then, with the $y \rightarrow 1$ limit of Eq.
(\ref{69}) holding for the general $Q_{\nu}^2(y)$ with arbitrary $\nu$, we
see that the real parts of the $Q_{\nu}^2(y)$ wave functions all have the
same $1/(y-1)$ leading behavior at $y = 1$ as the massless
graviton itself, with the non-normalizable modes all diverging at $y=1$
at one and the same rate.

In order to test for completeness in the ${\rm Re} [Q_{\nu}^2(y)]$ plus
massless graviton sector, we need to expand the localized square step
$\hat{V}_Q=\hat{V}$ when $|w_1|\leq |w|\leq |w_2|$, $\hat{V}_Q=0$
otherwise, in terms of these solutions as 
\begin{equation}
\hat{V}_Q= \sum_n V_{n}{\rm Re}[Q_{\nu}^2(y)]+\frac{V_0}{y^2-1}~~.
\label{76}
\end{equation}
(As previously, for clarity we use $n^2$ here to denote the squared masses
of the $Q_{\nu}^2(y)$ sector modes, and use $m^2$ for the $P_{\nu}^2(y)$
sector.) Given the asymptotic limit exhibited in Eq. (\ref{69}), in order
to cancel both the leading $1/(y-1)$ term and the next to leading $O(1)$
term from the right-hand side of Eq. (\ref{76}), we must constrain the
$V_n$ coefficients according to 
\begin{equation}
\sum_n V_{n}+\frac{V_0}{2}=0~~,~~\frac{1}{2}\sum_n
V_n(\nu^2+\nu-1)+\frac{V_0}{4}=0~~,
\label{77}
\end{equation}
to thus enable us to reexpress the square step
expansion as 
\begin{equation}
\hat{V}_Q= \sum_n V_{n}\left[{\rm
Re}[Q_{\nu}^2(y)]-\frac{2}{y^2-1}\right]~~,
\label{78}
\end{equation}
as subject to the constraint
\begin{equation}
\sum_n V_{n}\left[\nu^2+\nu-2\right]=-\sum_n
V_{n}\frac{n^2}{H^2}=0~~.
\label{79}
\end{equation}

On now applying
$\int_0^{\infty}d|w|e^{-2A}P_{\nu^{\prime}}^2(|w|)=(b/H^2)\int_1^{{\rm
coth}\sigma}dy P_{\nu^{\prime}}^2(y)$ to Eq. (\ref{78}) where
$\nu^{\prime 2}+\nu^{\prime}-2=-m^2/H^2$, use of the relations
\begin{equation}
\int_1^{{\rm coth}\sigma}dy P_{\nu^{\prime}}^2(y){\rm Re}[Q_{\nu}^2(y)]
=\frac{4H^2P(\nu^{\prime})}{(m^2-n^2)}~~,
\label{80}
\end{equation}
\begin{equation}
\int_1^{{\rm coth}\sigma}dy \frac{P_{\nu^{\prime}}^2(y)}{(y^2-1)}
=\frac{2H^2P(\nu^{\prime})}{m^2}~~,
\label{81}
\end{equation}
which follow from Eqs. (\ref{69}) and
(\ref{72}) (with $P(\nu^{\prime})=\nu^{\prime}(\nu^{\prime
2}-1)(\nu^{\prime}+2)/4$ now being given by $m^2(m^2-2H^2)/4H^4$) then
yields 
\begin{equation}
4b\sum_nV_nP(\nu^{\prime})
\left[\frac{1}{(m^2-n^2)} -\frac{1}{m^2}\right]=
\sum_nV_n 
{ b(m^2-2H^2)n^2\over H^4(m^2-n^2)}=B_m~~,
\label{82}
\end{equation}
where $B_m$ is the same function that was already given earlier in Eq.
(\ref{75}). 

\begin{figure}
\centerline{\epsfig{file=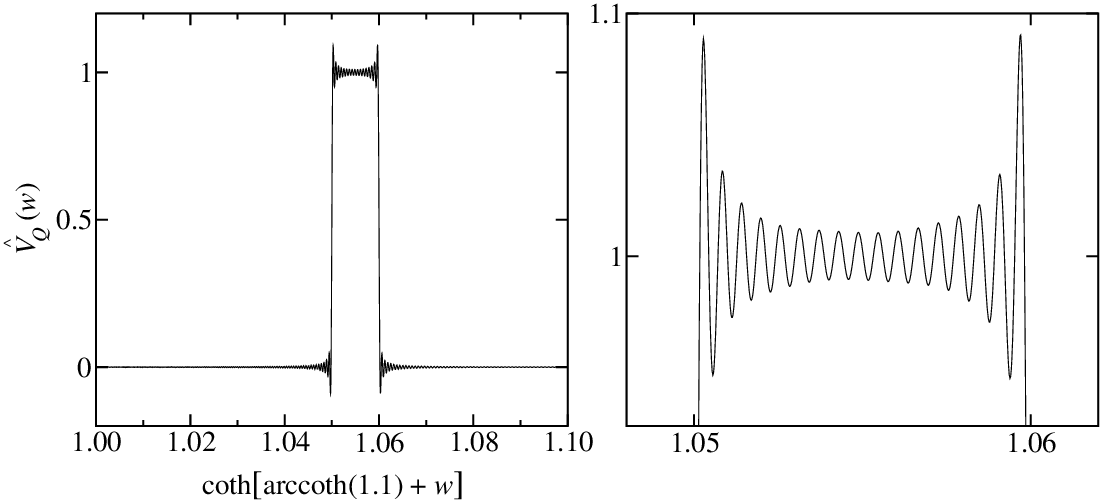,height=3.5in}}
\medskip
\caption{The left panel shows a reconstruction of the square step
 $\hat{V}_Q(w)=1$,  $1.05 \leq
{\rm coth} ({\rm arccoth}(1.1) +b|w|) \leq 1.06$, $\hat{V}_Q(w)=0$
otherwise, via the $dS_4^-$ discrete ${\rm Re}[Q_{\nu}^2({\rm coth}
(\sigma +b|w|))]$ plus ${\rm sinh}^2({\rm coth} (\sigma
+b|w|))$ mode basis in the typical case where ${\rm
coth}\sigma =1.1$,
$H/b=0.458$, and $b=1$. The right panel shows a blow-up of the region
near the top of the step.}
\label{Fig. (6)}
\end{figure}

Given Eq. (\ref{82}), the $V_n$ coefficients can now be found numerically,
and lead, for the case of the first 500 modes in the basis to the plot
displayed in Fig. (6) (i.e. we restrict to the first 500
$P^{1}_{\nu^{\prime}}({\rm coth}\sigma)$ zeroes and the first 500
${\rm Re}[Q^{1}_{\nu}({\rm coth}\sigma)]$ zeroes). As Fig. (6) thus
indicates, the divergent mode basis is every bit as capable of
reconstructing the square step as the convergent one and every bit as
capable of recovering the Gibbs phenomenon, and is thus every bit as
complete. As with our earlier examples then, we once again confirm that
completeness is not at all tied to normalizability.

\section{Final Comments}

In this work we have shown that in and of itself the requirement of
normalizability of basis modes is not at all needed for completeness, and
that one can construct localized steps out of bases whose modes are not
normalizable at all. Since the localized steps that we have constructed
out of non-normalizable bases involve expansion coefficients $V_n$
which are explicitly found to be finite, this suggests that we should be
able to construct propagators involving the modes in which these modes
appear as poles which have residues which are themselves finite. Thus
in sharp contrast to the situation in which propagators are built
out of normalizable modes, for propagators which are built out modes of
which are not normalizable, these residues must then not be related
to normalization constants or to any bilinear integrals of the modes
at all for that matter. 

To explicitly construct such divergent mode based propagators we must
first introduce explicit source terms. For the case of interest to the
brane world the source is typically taken to be a transverse-traceless
energy-momentum tensor $S^{^{TT}}_{\mu\nu}$ which is confined to the brane
at $w=0$, with Eqs. (\ref{10}) and (\ref{11}) being replaced by (see e.g.
\cite{Mannheim2005})
\begin{equation}
\left[{\partial^2 \over \partial |w|^2}
-4\left({d A\over
d|w|}\right)^2
+e^{-2A}\tilde{\nabla}_{\alpha}\tilde{\nabla}^{\alpha}
\right]h^{^{TT}}_{\mu\nu} =0~~,
\label{83}
\end{equation}
\begin{equation}
\delta(w)\left[{\partial \over
\partial |w|}-2{d A\over
d|w|}\right]h^{^{TT}}_{\mu\nu}=-\kappa_5^2\delta(w)S^{^{TT}}_{\mu\nu}~~,
\label{84}
\end{equation}
where $\kappa_5^2$ is the brane-world gravitational constant.

For the case first of the convergent warp factor $M_4^+$ brane world where
Eqs. (\ref{83}) and (\ref{84}) reduce to 
\begin{equation}
\left[{\partial^2 \over \partial |w|^2}
-4b^2
+e^{2b|w|}\eta^{\alpha\beta}\partial_{\alpha}\partial_{\beta} \right]
h^{^{TT}}_{\mu\nu} =0~~,
\label{85}
\end{equation}
\begin{equation}
\delta(w)\left[{\partial \over
\partial|w|}+2b\right]h^{^{TT}}_{\mu\nu}=
-\kappa_5^2\delta(w)S^{^{TT}}_{\mu\nu}~~,
\label{86}
\end{equation}
on recalling that the Bessel functions obey
\begin{equation}
\left[{d \over d|w|}+2b\right]\left[\alpha_qJ_2\left({qe^{b|w|}
\over b}\right)+\beta_q Y_2\left({qe^{b|w|}
\over b}\right)\right]=qe^{b|w|}\left[\alpha_qJ_1\left({qe^{b|w|}
\over b}\right)+\beta_q Y_1\left({qe^{b|w|}
\over b}\right)\right]~~,
\label{87}
\end{equation}
an explicit solution to Eqs. (\ref{85}) and (\ref{86})  can readily be
given,  viz.
\cite{Giddings2000}  
\begin{eqnarray}
h^{^{TT}}_{\mu\nu}(x,|w|)&&=-{\kappa_5^2\over (2\pi)^4}\int
d^4x^{\prime}d^4pe^{ip\cdot(x- x^{\prime})}
{[\alpha_q
J_2(qe^{b|w|}/b)  +\beta_q
Y_2(qe^{b|w|}/b)]
 \over q[\alpha_q J_1(q/b)+\beta_q
Y_1(q/b)]}S^{^{TT}}_{\mu\nu}(x^{\prime})
\nonumber \\
&&=-2\kappa_5^2\int d^4x^{\prime}\hat{G}^{^{TT}}(x,
x^{\prime},w,0;\alpha_q,\beta_q,M_4^+) S^{^{TT}}_{\mu\nu}(x^{\prime})~~,
\label{88}
\end{eqnarray}
where $q^2=(p^0)^2-\bar{p}^2$ ($q$ being understood to have the same sign
as $p^0$ here), and $\alpha_q$ and $\beta_q$ are arbitrary constants. 

The generalization of this solution to the 
divergent warp factor $M_4^-$ brane world where we have
\begin{equation}
\left[{\partial^2 \over \partial |w|^2}
-4b^2
+e^{-2b|w|}\eta^{\alpha\beta}\partial_{\alpha}\partial_{\beta} \right]
h^{^{TT}}_{\mu\nu} =0~~,
\label{89}
\end{equation}
\begin{equation}
\delta(w)\left[{\partial \over
\partial|w|}-2b\right]h^{^{TT}}_{\mu\nu}=
-\kappa_5^2\delta(w)S^{^{TT}}_{\mu\nu}~~,
\label{90}
\end{equation}
and
\begin{eqnarray}
\left[{d \over
d|w|}-2b\right]&&\left[\alpha_qJ_2\left({qe^{-b|w|}
\over b}\right)+\beta_q Y_2\left({qe^{-b|w|}
\over b}\right)\right]
\nonumber \\
&&=-qe^{-b|w|}\left[\alpha_qJ_1\left({qe^{-b|w|}
\over b}\right)+\beta_q Y_1\left({qe^{-b|w|}
\over b}\right)\right]~~,
\label{91}
\end{eqnarray}
is of the form \cite{Mannheim2005}
\begin{eqnarray}
h^{^{TT}}_{\mu\nu}(x,|w|)&&={\kappa_5^2\over (2\pi)^4}\int
d^4x^{\prime}d^4pe^{ip\cdot(x- x^{\prime})}
{[\alpha_q
J_2(qe^{-b|w|}/b)  +\beta_q
Y_2(qe^{-b|w|}/b)]
 \over q[\alpha_q J_1(q/b)+\beta_q
Y_1(q/b)]}S^{^{TT}}_{\mu\nu}(x^{\prime})
\nonumber \\
&&=-2\kappa_5^2\int d^4x^{\prime}\hat{G}^{^{TT}}(x,
x^{\prime},w,0;\alpha_q,\beta_q,M_4^-) S^{^{TT}}_{\mu\nu}(x^{\prime})~~,
\label{92}
\end{eqnarray}
with $\alpha_q$ and $\beta_q$ again being arbitrary constants.
That the solution of Eq. (\ref{92}) satisfies Eq. (\ref{89}) follows
directly, since both $J_2(qe^{-b|w|}/b)$ and  $Y_2(qe^{-b|w|}/b)$
separately satisfy the Bessel function equation given as Eq. (\ref{19})
with $y$ being given by $y=qe^{-b|w|}/b$; and that the solution satisfies
Eq. (\ref{90}) follows from Eq. (\ref{91}). For this solution we note that
it is the requirement that Eq. (\ref{92}) obey Eq. (\ref{90}) (technically
the Israel junction condition in the presence of the source) which fixes
the overall normalization of the integrand in Eq. (\ref{92}), with none
of the $\alpha_q$ or $\beta_q$ coefficients needing to be infinite. In
fact the same is true of the $M_4^+$ brane world propagator as its
overall normalization is fixed by the junction condition of Eq.
(\ref{86}), with the similarity of the $M_4^+$ solution of Eq. ({88}) and
the $M_4^-$ solution of Eq. (\ref{92}) essentially showing complete
insensitivity to the normalizability or lack thereof of basis modes.

In order to be able to make contact with the various bases we used in our
construction of localized steps in the divergent warp factor $M_4^-$ brane
world, we need to make specific choices for the $\alpha_q$ and $\beta_q$
coefficients which appear in Eq. (\ref{92}). To make contact with the
convergent
$J_2(qe^{-b|w|}/b)$ modes, we recall that a Taylor series expansion of
$J_1(q/b)$ around any $j_i$ zero of $J_1$ is of the form  
\begin{equation}
J_1\left(q/b\right)=\left({q\over
b}-j_i\right)J_1^{\prime}(j_i)=\left({q\over b}-j_i\right)\left[{J_1(j_i)
\over j_i}-J_2(j_i)\right]=-\left({q\over b}-j_i\right)J_2(j_i)~~.
\label{93}
\end{equation}
Thus on setting $\alpha_q=1$, $\beta_q=0$ and recalling that each $j_i$
zero of $J_1(j_i)$ is also a zero of $J_1(-j_i)$, we see that the
propagator of Eq. (\ref {92}) contains a set of isolated poles at the
zeroes of $J_1$ (a pole at $q=bj_1$ when $p^0$ is positive, and a pole at
$q=-bj_1$ when $p^0$ is negative), with a $p^0$ plane contour integration
yielding a net pole contribution to the propagator of the form
\begin{equation}
\hat{G}^{^{TT}}(x,0,w,0;\alpha_q=1,\beta_q=0,M_4^-)=-i\sum_if_i(|w|)f_i(0)\int{ 
d^3p\over (2\pi)^3}
 {e^{i\bar{p}\cdot \bar{x}}
\over
2E_i}\left[e^{-iE_it}
-e^{iE_it}\right]~~,
\label{94}
\end{equation}
where 
\begin{equation}
f_i(|w|)={b^{1/2}J_2(j_ie^{-b|w|})\over
J_2(j_i)}~~,~~E_i=(\bar{p}^2+b^2j_i^2)^{1/2}~~,
\label{95}
\end{equation}
and where the summation in Eq. (\ref{94}) only needs extend over the
$j_i>0$ modes. Finally, recalling Eq. (\ref{32}), viz. 
\begin{equation}
\int_0^{\infty} d|w| e^{-2b|w|}J_2^2(me^{-b|w|}/b)
={J_2^2(m/b) \over 2b}~~,
\label{96}
\end{equation}
we see that the $f_i(|w|)$ basis modes precisely obey Eqs. (\ref{1})  and
(\ref{5}), with the pole structure of the $M_4^-$ brane-world propagator
$\hat{G}^{^{TT}}(x,0,w,0;\alpha_q=1,\beta_q=0,M_4^-)$ nicely recovering
the orthonormality and closure structure of the $J_2^2(me^{-b|w|}/b)$
sector basis modes.

In order to make contact with the non-normalizable $M_4^-$ mode
sector, we need to take $\beta_q$ to be  non-zero in the $M_4^-$
propagator. Recalling that
$J_{1}(y)$, $J_{2}(y)$, $Y_{1}(y)$ and $Y_{2}(y)$ respectively behave as
$y/2$, $y^2/8$, $-2/\pi y+O(y)$ and $-4/\pi y^2-1/\pi$ near $y=0$, we see
that once $\beta_q$ is non-zero, the integrand
$[\alpha_qJ_2(qe^{-b|w|}/b)  +\beta_q
Y_2(qe^{-b|w|}/b)]/q[\alpha_qJ_1(q/b)+\beta_q Y_1(q/b)]$ will behave as
$2be^{2b|w|}/q^2$ near $q^2=0$ independent of the actual values of
$\alpha_q$ and $\beta_q$, to thus give rise to a massless graviton pole
term contribution of the form 
\begin{equation}
\hat{G}^{^{TT}}(x,0,w,0;\alpha_q,\beta_q \neq 0,M_4^-,{\rm graviton})=
ibe^{2b|w|}\int {d^3p\over (2\pi)^3}
 {e^{i\bar{p}\cdot \bar{x}}
\over 2|p|}\left[e^{-i|p|t}-e^{i|p|t}\right]~~.
\label{97}
\end{equation}
Non-normalizable as the $M_4^-$ brane-world graviton might be, as we
see, it nonetheless appears in the propagator with a finite
residue \cite{footnote15}.

To make contact with the $M_4^-$ brane world divergent $Y_2(qe^{-b|w|}/b)$
modes we set $\alpha_q=0$ in
$\hat{G}^{^{TT}}(x,0,w,0;\alpha_q,\beta_q,M_4^-)$, and while we
immediately then obtain poles at the zeroes of $Y_1(q/b)$, since both
$Y_2(qe^{-b|w|}/b)$ and $Y_1(q/b)$ have branch points at
$q=0$, we also obtain a cut discontinuity, with the full singular term
evaluating to  \cite{Mannheim2005}
\begin{eqnarray}
&&\hat{G}^{^{TT}}(x,0,w,0;\alpha_q=0,\beta_q \neq
0,M_4^-)=ibe^{2b|w|}\int{d^3p 
\over (2\pi)^3} {e^{i\bar{p}\cdot \bar{x}}
\over 2|p|}\left[e^{-i|p|t}-e^{i|p|t}\right]
\nonumber \\
&&\phantom{=}-i\sum_i\tilde{f}_i(|w|)\tilde{f}_i(0)\int{  d^3p\over
(2\pi)^3}
 {e^{i\bar{p}\cdot \bar{x}}
\over
2E_i}\left[e^{-iE_it}
-e^{iE_it}\right]
\nonumber \\
&&\phantom{=}+{i \over
(2\pi)^3}\int d^3p {e^{i\bar{p}\cdot \bar{x}}
\over
2E_p}\left[e^{-iE_pt}
-e^{iE_pt}\right]\int  dm \left[1-2i{J_2(me^{-b|w|}/b) \over
Y_1(m/b)}\right]
\nonumber \\
&&\phantom{=}\times 
\left[{[Y_1(m/b)J_2(me^{-b|w|}/b)-J_1(m/b)Y_2(me^{-b|w|}/b)]\over
\pi [4J_1^2(m/b)+Y_1^2(mb)]} \right]~,
\nonumber \\
&&
\label{98}
\end{eqnarray}
where
\begin{equation}
\tilde{f}_i(|w|)={b^{1/2}Y_2(y_ie^{-b|w|})\over
Y_2(y_i)}~~,~~E_i=(\bar{p}^2+b^2y_i^2)^{1/2}~~.
\label{99}
\end{equation}
As we again see, despite the lack of normalizability of
$Y_2(me^{-b|w|}/b)$ modes, all the terms which appear in Eq. (\ref{98})
do so with coefficients which are nonetheless finite.

Further examples of this phenomenon may be found in the other divergent
warp factor brane worlds we have been considering. However, unlike the
exact propagator solutions of Eqs. (\ref{88}) and (\ref{92}), for the
$AdS_4$ and $dS_4$ based brane worlds so far such propagators have only
been constructed in low order. Specifically, for the $AdS_4^+$ brane
world for instance where the background metric of Eq. (\ref{6}) takes the
explicit form 
\begin{equation}
ds^2=dw^2+e^{2A(|w|)}[dx^2+e^{2Hx}(dy^2+dz^2-dt^2)]~
\label{100}
\end{equation}
with the $AdS_4^+$ warp factor $A(|w|)$ being given in Eq. (\ref{9}) and
$\lambda$ being positive, to lowest order in $H$ the appropriate $AdS_4^+$
propagator is given as \cite{Mannheim2005}
\begin{eqnarray}
&&\hat{G}^{^{TT}}(x,x^{\prime},w,0;
\hat{\alpha}_{\nu},\hat{\beta}_{\nu},AdS_4^+) 
\nonumber \\
&&\phantom{=}={1\over
2H(2\pi)^3}\int_{-\infty}^{\infty} dp^0dp^2dp^3\int_{0}^{\infty}dp^1p^1
B_{\nu}({\rm tanh}(b|w|-\sigma),\hat{\alpha}_{\nu},\hat{\beta}_{\nu})
e^{Hx/2}e^{Hx^{\prime}/2}~~~
\nonumber \\
&&\phantom{=}\phantom{=}\times 
e^{-ip^0(t-t^{\prime})+ip^2(y-y^{\prime})
+ip^3(z-z^{\prime})}J_{\tau}(ke^{-Hx}/H)
J_{\tau}(ke^{-Hx^{\prime}}/H)
\label{101}
\end{eqnarray}
where $k$ is given by $k=[(p^0)^2-(p^2)^2-(p^3)^2]^{1/2}$,
$\tau$ and $\nu$ are given by $\tau=\nu
+1/2=[9/4+k^2/H^2-(p^1)^2/H^2]^{1/2}$, and the quantity $B_{\nu}({\rm
tanh}(b|w|-\sigma),\hat{\alpha}_{\nu},\hat{\beta}_{\nu})$ is given by  
\begin{eqnarray}
&&B_{\nu}({\rm
tanh}(b|w|-\sigma),\hat{\alpha}_{\nu},\hat{\beta}_{\nu})~~~~~~
\nonumber \\
&&=\frac{1}{H(\nu -1)(\nu
+2)}\left[\frac{\hat{\alpha}_{\nu}
P_{\nu}^2({\rm tanh}(b|w|-\sigma))  +\hat{\beta}_{\nu}
Q_{\nu}^2({\rm tanh}(b|w|-\sigma))}{\hat{\alpha}_{\nu} P_{\nu}^1(-{\rm
tanh}\sigma)+\hat{\beta}_{\nu}
Q_{\nu}^1(-{\rm tanh}\sigma)}
\right]~~.
\label{102}
\end{eqnarray}
As constructed the quantity $B_{\nu}({\rm
tanh}(b|w|-\sigma),\hat{\alpha}_{\nu},\hat{\beta}_{\nu})$ obeys 
\begin{equation}
\delta(w)\left[\frac{d}{d|w|}-2\frac{dA}{d|w|}\right]B_{\nu}({\rm
tanh}(b|w|-\sigma),\hat{\alpha}_{\nu},\hat{\beta}_{\nu}) =\delta(w)~~,
\label{103}
\end{equation}
and has a small $H$ limit into the analogous $M_4^+$ integrand, viz.
\begin{equation}
B_{\nu}({\rm
tanh}(b|w|-\sigma),\hat{\alpha}_{\nu},\hat{\beta}_{\nu}) \rightarrow
{[\alpha_q J_2(qe^{b|w|}/b)  +\beta_q Y_2(qe^{b|w|}/b)]
 \over q[\alpha_q J_1(q/b)+\beta_q Y_1(q/b)]}~~,
\label{104}
\end{equation}
where $\hat{\alpha}_{\nu}=\alpha_{q}{\rm cos}(\nu\pi)+\beta_{q}{\rm
sin}(\nu\pi)$,
$\hat{\beta}_{\nu}=(2/\pi)[-\alpha_{q}{\rm sin}(\nu\pi)+\beta_{q}{\rm
cos}(\nu\pi)]$.
In the small $H$ limit $\hat{G}^{^{TT}}(x,x^{\prime},w,0;
\hat{\alpha}_{\nu},\hat{\beta}_{\nu},AdS_4^+)$ obeys 
\begin{eqnarray}
&&\left[{\partial^2 \over \partial w^2}
-4\left({dA \over d|w|}\right)^2-4 {dA \over d|w|}\delta(w)
+e^{-2A}\tilde{\nabla}_{\alpha}\tilde{\nabla}^{\alpha}
\right]\hat{G}^{^{TT}}(x,x^{\prime},w,0;
\hat{\alpha}_{\nu},\hat{\beta}_{\nu},AdS_4^+)
\nonumber \\
&&\phantom{=}=e^{Hx}\delta(x-x^{\prime})
\delta(t-t^{\prime})\delta(y-y^{\prime})\delta(z-z^{\prime})\delta(w)~~,
\label{105}
\end{eqnarray}
with the fluctuation
\begin{equation}
h_{\mu\nu}^{^{TT}}(x,|w|)=-2\kappa_5^2\int d^4x^{\prime}e^{-Hx^{\prime}}
\hat{G}^{^{TT}}(x,x^{\prime},w,0;
\hat{\alpha}_{\nu},\hat{\beta}_{\nu},AdS_4^+)
S_{\mu\nu}^{^{TT}}(x^{\prime}) 
\label{106}
\end{equation}
thus being an exact $AdS_4^+$ brane world small $H$ solution to the 
$AdS_4^+$ variant of Eqs. (\ref{83}) and (\ref{84}) for an arbitrary
$S_{\mu\nu}^{^{TT}}(x^{\prime})$ source on the brane.

As regards pole terms in the $AdS_4^+$ brane-world propagator, since
$(\nu-1)(\nu+2)=q^2/H^2$, the
$(\nu-1)(\nu+2)$ term in $B_{\nu}({\rm
tanh}(b|w|-\sigma),\hat{\alpha}_{\nu},\hat{\beta}_{\nu})$ generates a 
massless $\nu=1$ graviton pole in the propagator which is found to be of
the form \cite{Mannheim2005} 
\begin{eqnarray}
&&\hat{G}^{^{TT}}(x,x^{\prime},w,0;\hat{\alpha}_{\nu},
\hat{\beta}_{\nu},AdS_4^+, {\rm graviton})
\nonumber \\
&&~~=\frac{be^{2A}
\hat{D}_S(x,x^{\prime},m=0)}{
[-(\hat{\alpha}_1/\hat{\beta}_1)(H^2/b^2)+(1-H^2/b^2)^{1/2}+(H^2/b^2){\rm
arccosh}(b/H)]}~~,~~~~~
\label{107}
\end{eqnarray}
where $\hat{D}_S(x,x^{\prime},m)$ is the pure $AdS_4$ spacetime
propagator which obeys 
\begin{equation}
[\partial_x^2-H\partial_x+e^{-2Hx}(\partial_y^2+\partial_z^2-\partial_t^2)
-2H^2-m^2]
\hat{D}_S(x,x^{\prime},m)=e^{Hx}\delta^4(x-x^{\prime})~~.
\label{108}
\end{equation}
As we see, despite the lack of normalizability of the graviton wave
function, the residue at the $AdS_4^+$ massless graviton pole is
nonetheless finite \cite{footnote16}. Similarly, if we set
$\hat{\alpha}_{\nu}=0$ in Eq. (\ref{101}) we will immediately generate
the divergent $Q_{\nu}^2({\rm tanh}(b|w|-\sigma))$ modes as poles
associated with the zeroes of
$Q_{\nu}^1(-{\rm tanh}\sigma)$, with these pole terms also possessing
finite residues. Consequently, in the brane world divergent modes are
fully capable of appearing with finite residues in propagators and their
lack of normalizability should not be taken as being a criterion for
excluding them. In fact, with the $M_4^+$ propagator $\hat{G}^{^{TT}}(x,
x^{\prime},w,0;\alpha_q,\beta_q,M_4^+)$ of Eq. (\ref{88})
being causal when we set $\alpha_q=1$, $\beta_q=i$
\cite{Giddings2000,Mannheim2006} (so that it is then
based on outgoing Hankel functions), given the small $H$ limit of
$B_{\nu}({\rm tanh}(b|w|-\sigma),\hat{\alpha}_{\nu},\hat{\beta}_{\nu})$
exhibited in Eq. (\ref{104}), it will be the
$\hat{G}^{^{TT}}(x,x^{\prime},w,0;\hat{\alpha}_{\nu},
\hat{\beta}_{\nu},AdS_4^+)$ propagator with
$\hat{\alpha}_{\nu}=e^{i\pi\nu}$,
$\hat{\beta}_{\nu}=(2i/\pi)e^{i\pi\nu}$ which will be the $AdS_4^+$ analog
of the outgoing Hankel function based causal $M_4^+$ brane-world
propagator, with this particular $AdS_4^+$ brane world propagator
explicitly being found to be causal \cite{Mannheim2005}. As such, the
causal $AdS_4^+$ brane-world propagator with 
$\hat{\alpha}_{\nu}=e^{i\pi\nu}$,
$\hat{\beta}_{\nu}=(2i/\pi)e^{i\pi\nu}$ possesses an explicit massless
graviton pole whose residue is finite, with there thus being no
justification for excluding it \cite{footnote17}.

\begin{acknowledgments}
The authors would like to thank Dr. A. H. Guth, Dr. D. I. Kaiser and Dr.
A. Nayeri for their active participation in this work, and for
their many helpful comments.
\end{acknowledgments}


\begin{thebibliography}{}
\bibitem{footnote1} Even in quantum mechanics one does not discard plane
wave modes even though they cause the integral on the right-hand side of
Eq. (\ref{18}) to diverge, since divergent as they may be, one can still
construct localized wave packets out of them. In this respect then, the
point of this paper will be to construct localized configurations out of
basis vectors which diverge even more rapidly than plane waves. And while
we shall restrict the study of this paper to the classical-mechanical
context, we note that within a quantum-mechanical context such localized
configurations could still belong to a Hilbert space even if the basis
vectors themselves out of which they are built do not.
\bibitem{footnote2} While Eq. (\ref{33}) is given in closed form, the
actual sum over modes is itself done numerically.
\bibitem{footnote4} The zeroes of $J_1(y)$ and $Y_1(y)$ are simple,
discrete ones which interlace each other, with first three positive zeroes
of $J_1(y)$ for instance occurring at $3.832$, $7.016$ and $10.173$, and
with the $n$-th positive zero being well-approximated by $j_n\approx
(n+1/4)\pi$ when $n$ is large; while the first three positive zeroes of
$Y_1(y)$ occur at $2.197$, $5.430$ and $8.596$, with the $n$-th positive
zero being  well-approximated by $y_n\approx (n-1/4)\pi$ when $n$ is
large. While these particular approximations do not hold at small $n$, the
$n$ which appears in the $j_n\approx (n+1/4)\pi$ and $y_n\approx
(n-1/4)\pi$ expressions does denote the number of the zero (counting the
first positive zero as $n=1$), so that for $n$ large or small  these
expressions give a correct counting of the number of zeroes.
\bibitem{footnote5} The reconstruction of the square step using the
divergent mode basis is so good that the only perceptible difference
between Figs. (1) and (2) is that in the region close to $e^{-w}=0$ the
$J_2(me^{-b|w|}/b)$ contribution is ever so slightly thicker. (The
constraints of Eq. (\ref{37}) force a more rapid convergence on the
$Y_2(ne^{-b|w|}/b)$ mode sum.)
\bibitem{footnote6} For the $M_4^-$ brane world this is just
as well, since it could otherwise not contain any massless graviton.
\bibitem{footnote7} The typical case of ${\rm tanh}\sigma=0.9$ (viz.
$H/b=0.436$) yields $\nu= 1.088$, $2.216$ and $3.362$ as the three
lowest positive solutions to $P_{\nu}^1(-{\rm tanh}\sigma)=0$, with the 
$n$-th positive zero being well-approximated by $\nu_n\approx
(n+1/4)\pi/{\rm arccos}(-{\rm tanh}\sigma) -1/2$ when $n$ is large.
\bibitem{footnote8} It is possible that this might perhaps be the first
time that the Gibbs phenomenon has explicitly been demonstrated for
associated Legendre functions, and especially for the divergent
$Q_{\nu}^2(y)$ modes which we show below.
\bibitem{footnote9} The typical case of ${\rm tanh}\sigma=0.9$ yields
$\nu= 0.536$, $1.649$ and $2.788$ as the three lowest positive solutions
to $Q_{\nu}^1(-{\rm tanh}\sigma)=0$, with the $n$-th positive zero being
well-approximated by $\nu_n\approx (n-1/4)\pi/{\rm arccos}(-{\rm
tanh}\sigma) -1/2$ when $n$ is large, with the zeroes of $P_{\nu}^1(-{\rm
tanh}\sigma)=0$ and $Q_{\nu}^1(-{\rm tanh}\sigma)=0$ thus interlacing
each other. As regards the $Q_{\nu}^1(-{\rm tanh}\sigma)=0$ solutions, we
note additionally that the lowest positive one actually corresponds
to an $m^2<0$ tachyon since it has $\nu<1$.
\bibitem{footnote10} The construction is so good that the only perceptible
difference between Figs. (3) and (4) is that in the regions close to
the edges of the steps the Gibbs phenomenon overshoot as shown in the Fig.
(3) blow-up is ever so slightly closer to $1.1$ than the one shown in
the blow-up of Fig. (4).
\bibitem{footnote11} Since the negative tension $AdS_4^-$ brane world
with divergent warp factor $e^{A(|w|)}=H{\rm cosh}(\sigma+b|w|)/b$ also
has convergent $P_{\nu}^2(y)$ and divergent $Q_{\nu}^2(y)$ modes (where
now $y={\rm tanh}(b|w|+\sigma)$ with range ${\rm tanh}\sigma \leq y \leq
1$), its structure is analogous to that of the divergent warp factor
$AdS_4^+$ world, and so we do not seek completeness tests for it here.
\bibitem{footnote12}  With the arbitrary hypergeometric function
$F(a,b;c,z)$ being equal to one when its argument $z$ is taken to be zero,
the large $y$ limits of $P_{\nu}^{2}(y)$ and $Q_{\nu}^{2}(y)$ are readily
obtained from their $|y|>1$ hypergeometric function representations of the
form $P_{\nu}^{\mu}(y)=
2^{\nu+1}\Gamma(-2\nu-1)\Gamma^{-1}(-\nu)\Gamma^{-1}(-\nu-\mu)
(y+1)^{\mu/2-\nu-1}(y-1)^{-\mu/2}F(\nu+1,\nu-\mu+1;2\nu+2,2/(1+y))
+2^{-\nu}\Gamma(2\nu+1)\Gamma^{-1}(\nu+1)\Gamma^{-1}(\nu-\mu+1)
(y+1)^{\mu/2+\nu}(y-1)^{-\mu/2}F(-\nu,-\nu-\mu;-2\nu,2/(1+y))$, 
$Q_{\nu}^{\mu}(y)=e^{i\mu\pi}
2^{-\nu-1}\pi^{1/2}\Gamma(\nu+\mu+1)\Gamma^{-1}(\nu+3/2)
y^{-\nu-\mu-1}(y^2-1)^{\mu/2}
F(\nu/2+\mu/2+1,\nu/2+\mu/2+1/2;\nu+3/2,1/y^2)$. While these
representations show that $P_{\nu}^{\mu}(y)$ and $Q_{\nu}^{\mu}(y)$ will
in general be complex in the $|y|>1$ region, the form for
$P_{\nu}^{\mu}(y)$ shows that it will actually be real when $y$ and $\mu$
are real and the parameter $\nu$ takes the value $\nu=-1/2+i\lambda$ where
$\lambda$ is real, a value for which the quantity
$\nu(\nu+1)=(\nu+1/2)^2-1/4$ which appears in the defining equation for
the associated Legendre functions of Eq. (\ref{64}) is then given as the
real $\nu(\nu+1)=-\lambda^2-1/4$. With Eq. (\ref{64}) remaining real at
$\nu=-1/2+i\lambda$, for such values of $\nu$ the then real
$P_{\nu}^{2}(y)$ and the real and imaginary parts of $Q_{\nu}^{2}(y)$
will all separately obey it. However, since Eq. (\ref{64})
can only have two independent solutions, it must be the case that one of
these three classes of solutions is redundant. On noting that no matter
what the value of $\nu$, the divergent part of $Q_{\nu}^2(y)$ at $y=1$ is
real while $P_{\nu}^2(y)$ is well-behaved there, we thus anticipate that
when $y$ is real and greater than one, it must be the (thus well-behaved
at $y=1$) imaginary part of $Q_{-1/2+i\lambda}^2(y)$ which must coincide
with the real $P_{-1/2+i\lambda}^2(y)$; and since it is not
immediately obvious how one may explicitly check such a connection
analytically, we have instead confirmed it numerically. In the
following then we can restrict the discussion to the use of
$P_{-1/2+i\lambda}^2(y)$ and ${\rm Re}[Q_{-1/2+i\lambda}^2(y)]$ as basis
modes (in both the $dS_4^+$ and the $dS_4^-$ brane worlds). As well as
enabling us to show that $P_{\nu}^{\mu}(y)$ is real for real $y$, real
$\mu$ and complex $\nu=-1/2+i\lambda$, the above representations of the
$P_{\nu}^{\mu}(y)$ and $Q_{\nu}^{\mu}(y)$ are also of use for actual
computational purposes when $y$ is greater than one, since for argument
$|z|<1$ a hypergeometric function can be represented as the absolutely
convergent power series
$F(a,b;c,z)=[\Gamma(c)/\Gamma(a)\Gamma(b)]\sum_{n=0}^{\infty}
\Gamma(a+n)\Gamma(b+n)z^n/[\Gamma(c+n)n!]$. Moreover, for large
values of the parameter $\lambda$, the functions
$P_{-1/2+i\lambda}^{\mu}(y)$ and ${\rm Re}[Q_{-1/2+i\lambda}^{\mu}(y)]$
can even be approximated by
$P_{-1/2+i\lambda}^{\mu}({\rm cosh}\theta)
=\lambda^{\mu-1/2}(2/\pi {\rm
sinh}\theta)^{1/2}{\rm cos}(\lambda\theta+\mu\pi/2-\pi/4)
-\lambda^{\mu-3/2}(1/2\pi {\rm
sinh}\theta)^{1/2}(\mu-1/2)(\mu+1/2){\rm coth}\theta{\rm
sin}(\lambda\theta+\mu\pi/2-\pi/4)$ and
${\rm Re}[Q_{-1/2+i\lambda}^{\mu}({\rm
cosh}\theta)]
=\lambda^{\mu-1/2}(\pi/2 {\rm sinh}\theta)^{1/2}{\rm
cos}(\lambda\theta+\mu\pi/2+\pi/4) 
-\lambda^{\mu-3/2}(\pi/8{\rm
sinh}\theta)^{1/2}(\mu-1/2)(\mu+1/2){\rm
coth}\theta{\rm sin}(\lambda\theta+\mu\pi/2+\pi/4).$ (It is
necessary to carry the first non-leading terms here since the
oscillatory leading terms can vanish at some specific $\theta$ values.)
\bibitem{footnote13}
With the $dS_4^-$ brane world range for $y$ being restricted to the
finite range $1\leq y\leq {\rm coth}\sigma$, in cases in which we
restrict to ${\rm coth}\sigma <3$, we are actually able to use
an extremely compact representation for evaluation of $P_{\nu}^2(y)$,
$P_{\nu}^1(y)$ and $P_{\nu}(y)$, viz. the form $P_{\nu}^{m}(y)=
(y^2-1)^{m/2}\Gamma(\nu+m+1)
F(-\nu+m,\nu+m+1,m+1,(1-y)/2)/[2^mm!\Gamma(\nu-m+1))]$ which holds for
any positive integer $m$, and the limiting form
$P_{\nu}(y)=F(-\nu,\nu+1,1,(1-y)/2))$ which holds when $m=0$, as
each of these hypergeometric function representations can be written as
a power series which is absolutely convergent over the entire $1 \leq y
\leq 3$ range.  From these representation we find in a typical case with
${\rm coth}\sigma=1.1$ that the three lowest positive
$\lambda$ solutions to $P_{-1/2+i\lambda}^1({\rm coth}\sigma)=0$ are
given as
$\lambda= 8.624$, $15.808$ and $22.930$, with the $n$-th
positive solution being well-approximated by $\lambda_n\approx
(n+1/4)\pi/{\rm arccosh}({\rm coth}\sigma)$ when $n$ is large.
\bibitem{footnote14} The typical case of ${\rm coth}\sigma=1.1$ yields
$\lambda= 4.928$, $12.231$ and $19.373$ as the three lowest positive
$\lambda$ solutions to ${\rm Re}[Q_{-1/2+i\lambda}^1({\rm
coth}\sigma)]=0$, with the
$n$-th positive solution being well-approximated by $\lambda_n\approx
(n-1/4)\pi/{\rm arccosh}({\rm coth}\sigma)$ when $n$ is large.
\bibitem{footnote15} Despite the fact that the negative tension $M_4^-$
brane world possesses a massless graviton whose residue is finite, we
note that its residue appears with an overall minus sign (viz.
negative signature) compared to the otherwise identical in
structure positive signature massless graviton residue of the positive
tension
$M_4^+$ brane world (compare the first forms given for
$h_{\mu\nu}^{^{TT}}(x,|w|)$ given in Eqs. (\ref{88}) and (\ref{92}) which
differ by an overall minus sign occasioned by the overall difference
in sign between the right-hand sides of Eqs. (\ref{87}) and
(\ref{91})).  Such negative signature is thought to indicate an
instability of the $M_4^-$ brane world. Nonetheless, even though
the $M_4^-$ brane world might thus not be of direct physical interest, it
can still serve as a useful mathematical laboratory for exploring the
completeness properties of bases built out of non-normalizable modes.
\bibitem{footnote16} Unlike the massless graviton of the negative tension
$M_4^-$ brane world, the massless graviton of the positive tension
$AdS_4^+$ brane world has a residue with a perfectly acceptable positive
signature -- as indeed it must since its residue continues into that of
the positive signature massless graviton of the positive tension $M_4^+$
brane world in the limit in which $H$ is taken to zero.
\end{thebibliography}
\end{document}